\documentclass{emulateapj}
\pdfoutput=1
\usepackage{graphicx}
\usepackage{txfonts}

\shorttitle{Magnetized galactic wind in IC\,10}
\shortauthors{K.T. Chy\.zy et al.}

\begin{document}

\title{The magnetized galactic wind and synchrotron halo of the
  starburst dwarf galaxy IC\,10}

\author{Krzysztof T. Chy\.zy\altaffilmark{1}}
\author{Robert T. Drzazga\altaffilmark{1}}
\author{Rainer Beck\altaffilmark{2}}
\author{Marek Urbanik\altaffilmark{1}}
\author{Volker Heesen\altaffilmark{3}}
\author{Dominik J. Bomans\altaffilmark{4}}
\altaffiltext{1}{Astronomical Observatory, Jagiellonian University, 
ul. Orla 171, 30-244 Krak\'ow, Poland}
\altaffiltext{2}{Max-Planck-Institut f\"ur Radioastronomie, Auf dem
H\"ugel 69, 53121 Bonn, Germany}
\altaffiltext{3}{School of Physics \& Astronomy, University of Southampton,
Southampton SO17 1BJ, UK}
\altaffiltext{4}{Ruhr-Universit\"at Bochum, Universit\"atsstrasse 150,
44780 Bochum, Germany}

%\date{Received date/ Accepted date}

\begin{abstract}
We aim to explore whether strong magnetic fields can be effectively generated in low-mass dwarf galaxies and, if
so, whether such fields can be affected by galactic outflows and spread out into the intergalactic medium (IGM).
We performed a radio continuum polarimetry study of IC\,10, the nearest starbursting dwarf galaxy, using a
combination of multifrequency interferometric (VLA) and single-dish (Effelsberg) observations. VLA observations
at  $1.43$\,GHz reveal an extensive and almost spherical radio halo of IC\,10 in total intensity, extending twice more
than the infrared-emitting galactic disk. 
The halo is magnetized with a magnetic field strength of 7\,$\mu$G in
the outermost parts. 
Locally, the magnetic field reaches about $29\,\mu$G in \ion{H}{2} complexes, 
becomes more ordered, and weakens to $22\,\mu$G in the synchrotron superbubble and to 
7--$10\,\mu$G within \ion{H}{1} holes. At the higher frequency of $4.86$\,GHz, we found
a large-scale magnetic field structure of X-shaped morphology, similar to that observed 
in several edge-on spiral galaxies.  The X-shaped magnetic structure can be caused by the 
galactic wind, advecting magnetic fields injected into the interstellar medium by stellar 
winds and supernova explosions. The radio continuum scale heights at 1.43\,GHz indicate the 
bulk speed of cosmic-ray electrons outflowing from \ion{H}{2} complexes of about 
60\,km\,s$^{-1}$, exceeding the escape velocity of 40\,km\,s$^{-1}$. Hence, the magnetized 
galactic wind in IC\,10 inflates the extensive radio halo visible at 1.43\,GHz and can seed the 
IGM with both random and ordered magnetic fields. These are signatures 
of intense material feedback onto the IGM, expected to be prevalent in the 
protogalaxies of the early Universe.
\end{abstract}


\section{Introduction}
\label{s:intro}

The generation, evolution, and role of magnetic fields in low-mass galaxies are still poorly
understood. Massive  late-type spiral galaxies  are supposed to have suitable conditions 
to generate large-scale magnetic fields via the large-scale $\alpha$--$\Omega$ dynamo 
(Beck et al.\ \citeyear{beck96}). The interstellar medium (ISM) in these galaxies has 
sufficiently strong shearing motions, caused by a combination of differential rotation and 
the Coriolis force acting on turbulent gas motions.  Low-mass, dwarf irregular galaxies
have much more chaotic gas motions and slow rotation  speeds with little rotational shear, 
which  might be unsuitable for the generation of large-scale magnetic fields.
 The majority of massive spiral galaxies has spiral magnetic fields related to
spiral density waves (Beck \& Wielebinski \citeyear{beck13}). In dwarf galaxies, 
without any coherent density wave structures, the  magnetic field topology could 
potentially be  very different and  regulated by other processes.

In dwarfs with high star formation rates (SFRs), magnetic fields could be strongly modified
by stellar winds, which constitute the dominant feedback in galaxy formation
and evolution (Veilleux et al.\ \citeyear{veilleux05}).  Because low-mass objects have
shallow gravitational potentials, they  are easily subjected to large-scale (galactic) 
winds (Recchi \& Hensler \citeyear{recchi13}). They are therefore suspected to potentially spread
out magnetic fields and cosmic rays (CRs) far away from star-forming disks that may 
even pervade the intergalactic medium (IGM). Thus, they are among the best
candidates for the cosmic magnetizers in the early Universe, where primordial 
galaxies may resemble some nearby starbursting low-mass galaxies (Kronberg 
et~al.\ \citeyear{kronberg99}; Dubois \& Teyssier \citeyear{dubois10}). Still, we 
lack observational evidence that stellar feedback in starbursting dwarfs can indeed produce
powerful galactic winds that are able to shape the global magnetic field structure.

One of the most nearby dwarf irregular galaxies, the Local Group member IC\,10, is a very
favorable target to test our current  knowledge of the generation and evolution of
magnetic fields in low-mass objects. The galaxy has a small linear size of about 
$1.6$\,kpc in diameter\footnote{We use the optical extent of IC\,10 of $6.8\arcmin$ 
from NED and the distance of 830\,kpc from Sanna et~al.\ (\citeyear{sanna08})} 
and  a low mass,  for instance, the mass of neutral hydrogen  of 
$M_\mathrm{HI}=9.8 \times 10^{7}\,M_{\sun}$ is more than four times
lower than  that of the Small Magellanic Cloud (SMC; Grebel et al.\ \citeyear{grebel03}).
IC\,10 is remarkable for its high SFR and large number of \ion{H}{2}
regions (Hodge \& Lee 1990). Normalized to its H$_2$ surface density, IC\,10 has
a much higher rate of star formation than most observed spiral or dwarf galaxies
(Mateo \citeyear{mateo98}; Leroy et al.\ \citeyear{leroy06}). IC\,10 has  also a large population
of Wolf-Rayet (W-R) stars and the highest surface density of W-R stars
among all Local Group galaxies (Massey \& Armandroff \citeyear{massey95}).
A high number  of carbon-type W-R stars  indicates that the galaxy experienced a brief,
but intense, galaxy-wide burst of star formation within the last 10\,Myr.
The observed stellar population  holds clues that the galaxy experienced another
starburst in the past, $\approx$150--500\, Myr ago (Vacca et al.\ \citeyear{vacca07}).

This galaxy shows a complex \ion{H}{1} morphology and peculiar velocity field 
(Ashley at al.\ \citeyear{ashley14}). While the optical disk is aligned at
a position angle of $PA=125\degr$ and at an inclination angle of $i=31\degr$ 
(according to the HyperLeda database), the \ion{H}{1} disk has a different 
orientation: $PA\approx 75\degr$ and $i\approx 50\degr$ (Wilcots \& Miller \citeyear{wilcots98}).
It shows that the ISM in IC\,10 is highly disturbed and the galactic disk is
only poorly defined. It is possible that IC\,10 has merged with another dwarf galaxy and is accreting
intergalactic gas filaments, which are visible in \ion{H}{1} data cubes at
various radial velocities. Both processes could have fed IC\,10's disk and 
thereby triggered its current starburst.

IC\,10 has also been investigated in the radio domain. In a polarization 
study by Chy\.zy et~al.\ (\citeyear{chyzy03}), using high-frequency
($10.45$\,GHz), low-resolution (73$\arcsec$) single-dish observations, a single 
polarized region in the southern part of the galaxy was found, close to the 
position where Yang \& Skillman (\citeyear{yang93}) found a large nonthermal 
superbubble. The rest of the disk of IC\,10 was devoid of polarized signal. 
The studies of IC\,10 were recently followed up by the high-resolution C-band
JVLA observations of Heesen et~al.\ (\citeyear{heesen11}), which confirmed
strong polarized emission in the region of the nonthermal superbubble. It was suggested
that the ordered magnetic field can be enhanced by the gas compression of the
expanding bubble. The superbubble can be powered by about 10 supernova explosions
(Yang \& Skillman \citeyear{yang93}) or one hypernova  event (Lozinskaya \& Moiseev
\citeyear{lozinskaya07}).  It is only a few Myr old and has yet to sweep up a shell of
atomic hydrogen (Heesen et al.\ \citeyear{heesen15}). No other coherent polarized 
structures were found in IC\,10 so far.

In this paper, we investigate the structure, origin, and evolution of the magnetic 
field in IC\,10 utilizing sensitive multiband polarimetric observations obtained 
with the VLA and the 100-m Effelsberg telescope. Radio interferometers cannot 
detect emission at large angular scales, resulting in the so-called ``missing 
spacings flux,'' which can be corrected for with single-dish observations. 
For instance, Heesen et al.\ (\citeyear{heesen11}) found that in their
JVLA observations of IC\,10  at 6\,GHz the flux density was $\approx$30\% lower
compared with the single-dish data presented by Chy\.zy et~al.\ (\citeyear{chyzy03}).
The study presented here is based on  a combination of VLA and Effelsberg 
data at $4.86$ and $8.46$\,GHz, which allows us to study the largest
angular scales found in weak, diffuse radio emission. Together with the VLA 
observations at $1.43$\,GHz, we are able to trace the synchrotron emission and 
magnetic fields not only in the galaxy's disk but also in its halo.

In Sect.~2 we present our radio observations of IC\,10 and data reduction procedures.
We present our results and their analysis in Sect.~\ref{s:results},
followed by a discussion of the magnetic field structure and galactic wind
(Sect.~\ref{s:discussion}). In Sect.~\ref{s:conclusions} we summarize our conclusions.

\section{Observations and data reduction}

\subsection{Effelsberg}
\label{s:eff}

IC\,10 was observed with  the Effelsberg 100-m telescope\footnote{The 
100-m telescope at Effelsberg is operated by the Max-Planck-Institut f\"ur
Radioastronomie (MPIfR) on behalf of the Max-Planck-Gesellschaft.}
at $4.85$ and $8.35$\,GHz. At $4.85$\,GHz we used the dual-horn system (at 
an angular horn separation of 8\arcmin) and a correlation polarimeter in 
the backend. Corrections to the telescope pointing model were
made about every $1.5$\,hr by scanning a nearby strong  unresolved radio source.
The individual coverages were obtained in both azimuth and elevation directions.
In order to establish a flux density scale, we observed the calibration source 3C\,286 and
assumed its total flux as $7.47$\,Jy, according to the flux scale established
from VLA observations by Perley \& Butler \citeyear{perley13}).

The data were reduced in the NOD2 package (Haslam \citeyear{haslam74}), 
which allowed us to combine the coverages using software beam-switching (Emerson 
et al.\ \citeyear{emerson79}) and spatial-frequency weighting methods (Emerson \&
Gr\"ave~\citeyear{emerson88}). The maps were then digitally filtered to remove the
spatial frequencies corresponding to noisy structures smaller than the telescope beam 
of $2\farcm 5$. The procedure of data reduction included  masking of spurious emission of
individual coverages and combining them into the final maps in Stokes parameters I,Q and U.
Details of observations and the obtained rms noise levels are given in Table \ref{t:ic10eff}.

At $8.35$\,GHz the single-horn receiver in the secondary focus was used. High 
sensitivity of the observations was obtained by using a frequency bandwidth of $1.1$\,GHz.
The galaxy was  mapped employing scans along R.A.\ and Dec orientations. The calibrator 3C\,286 was
once again applied for the flux scale calibration, assuming $5.27$\,Jy total flux according
to the VLA flux scale. We then combined individual coverages through a procedure similar 
as for $4.85$\,GHz, obtaining the final maps in Stokes I, Q and U with a resolution of $1\farcm 4$.

The final maps from both frequencies were converted to {\small FITS} format and taken
into the  Astronomical Image Processing System ({\small AIPS})\footnote{{\scriptsize AIPS}, 
the Astronomical Image Processing Software, is free software available from the NRAO.}. 
Maps of the total-power radio continuum intensity (TP), linearly polarized intensity (PI),
degree of polarization, and E-vector orientation angles were thus obtained, and 
rms noise levels were determined (Table \ref{t:ic10eff}). The PI 
was corrected for the positive bias resulting from the Ricean distribution of noise.
If not stated otherwise in the text, we present the magnetic field orientation, 
projected on the sky plane, as apparent B-vectors, which are defined as the 
E-vectors rotated by $90\degr$,  not corrected for Faraday rotation.

\begin{table*}
\caption{Parameters of the  Effelsberg observations}
\begin{center}
\begin{tabular}{ccccccccc}
\hline
\hline
Obs. Freq.   & Obs.\ Date & No. of Col/Rows   & No. of               & S$_\mathrm{TP}$      & S$_\mathrm{PI}$     & rms\ (TP)    & rms (PI) \\
 GHz         &           & in Coverages      & Coverages            &   (mJy)         &    (mJy)       & (mJy beam$^{-1}$)       & (mJy beam$^{-1}$)  \\
\hline
   $4.85$      & 2001      & 51 $\times$ 41     &  10                & 222 $\pm$ 9   & $6.3\pm 0.4$  & $0.80$           & $0.14$        \\
   $8.35$      & 2003      & 33 $\times$ 33     &  22                & 183 $\pm$ 8   & $2.5\pm 0.5$  & $0.50$           & $0.072$       \\
\hline
\end{tabular}
\end{center}
\label{t:ic10eff}
\end{table*}

\begin{table*}
\caption{Parameters of the VLA observations}
\begin{center}
\begin{tabular}{cccccc}
\hline
\hline
Band & Obs.\ Date               & Configuration & Integr. Time & rms (TP) & rms (PI) \\
     &                         &               & (hr)          & (mJy beam$^{-1}$)    &  (mJy beam$^{-1}$)   \\
\hline
   L & 2004 Mar. 27 / 2004 Aug 15 & C/D           & $5.0/4.3$      & 28$^a$  & 15$^a$ \\
   C & 2001 Dec. 4,28,31        & D             & $24.2$         & 17$^b$ & $6.5^b$ \\
   X & 2003 Feb. 9,10,14 / 2003 Apr. 11,12,27 / 2004 Aug. 18 & D      & $37.1$        & 10$^b$ & $3.5^b$ \\
\hline
\end{tabular}
\end{center}
$^a$ - Final rms noise level for the concatenated data from C and D configurations.
$^b$ - rms noise levels for the merged VLA and Effelsberg data (see text).
\label{t:ic10vla}
%\end{narrow}
\end{table*}

\subsection{VLA}
\label{s:vla}

\begin{figure*}
\centering
\includegraphics[clip,width=0.49\textwidth]{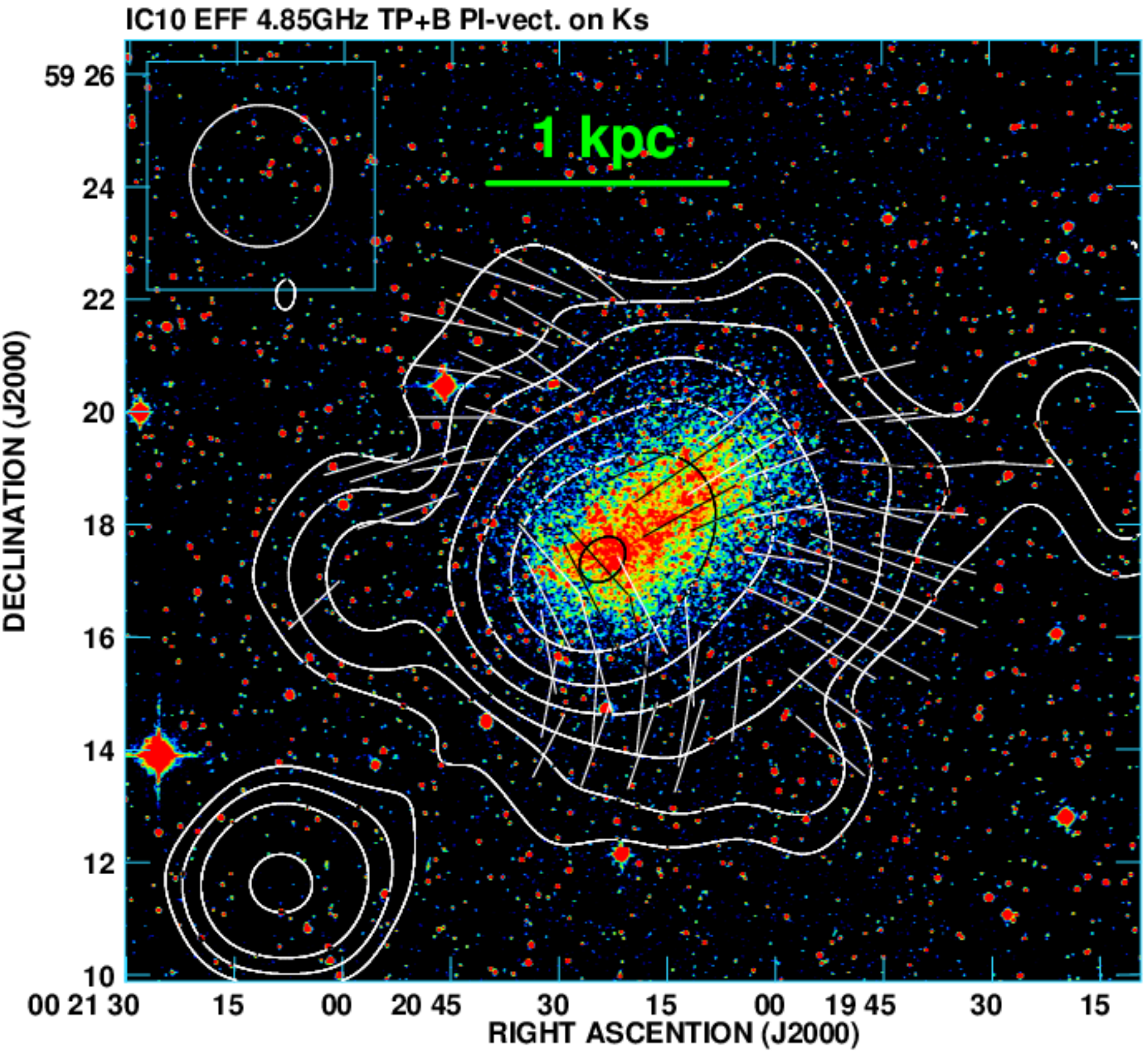}
\includegraphics[clip,width=0.49\textwidth]{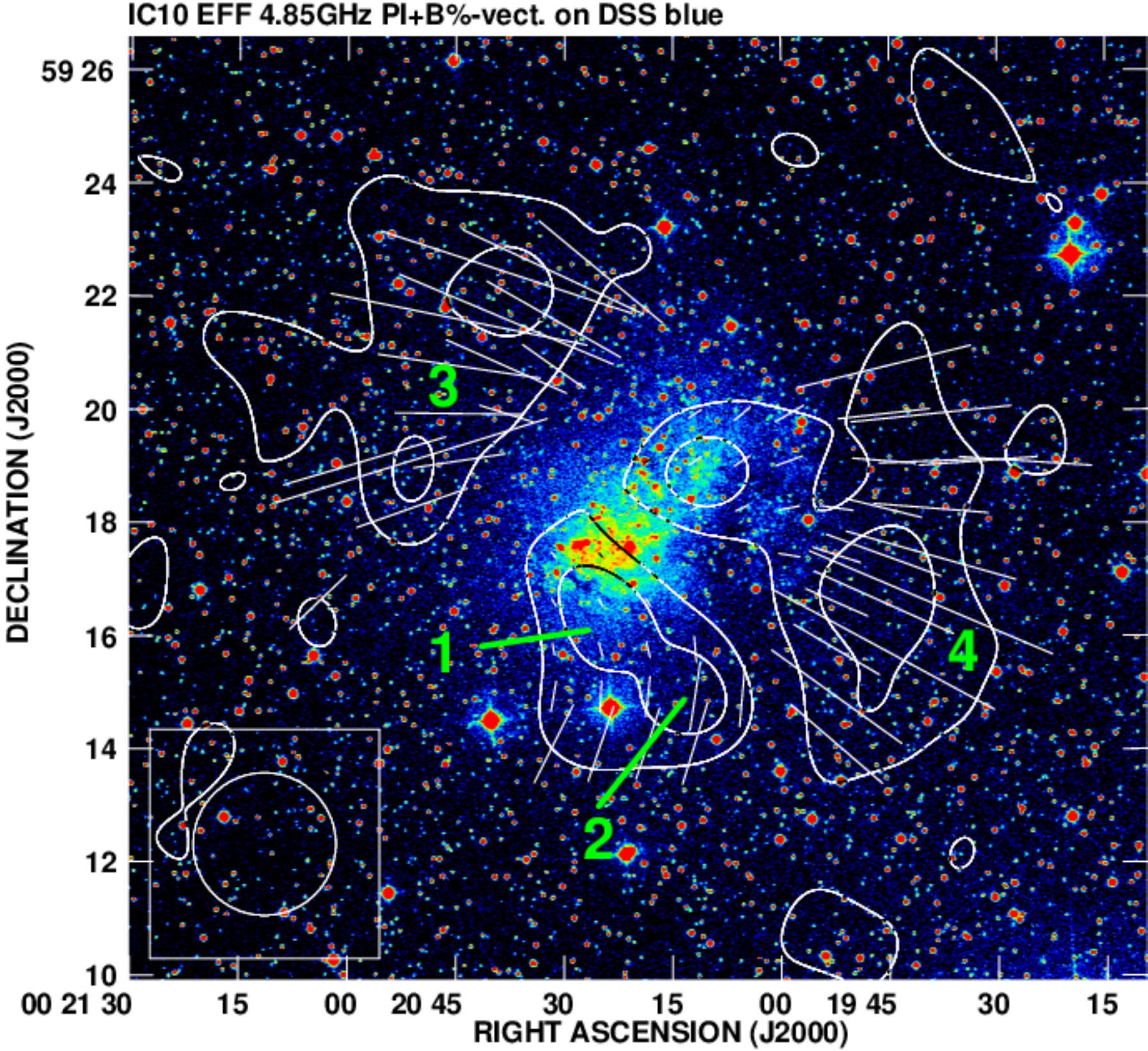}
\caption{Left: total-power contours and B-vectors of polarized intensity of IC\,10 at 4.85\,GHz
superimposed on the 2MASS Ks band ($2.2\,\mu$m) image. The contours levels are ($-3$, 3, 5, 8, 
16, 32, 64, 128) $\times$ 800\,$\mu$Jy\,beam$^{-1}$. A vector of $10\arcsec$ length corresponds to 
a polarized intensity of 59\,$\mu$Jy\,beam$^{-1}$. Right: Contours of polarized intensity and B-vectors 
of polarization degree of IC\,10 at $8.35$\,GHz superimposed on the DSS blue image. The contour 
levels are (3, 5) $\times$ 140\,$\mu$Jy\,beam$^{-1}$. A vector of $10\arcsec$ length corresponds to 
a polarization degree of 1\%. The resolution of the maps is 151$\arcsec$ $\times$ 151$\arcsec$ HPBW.}
\label{f:i10_6cm}
\end{figure*}

IC\,10 was observed with the VLA\footnote{The National Radio Astronomy Observatory is a 
facility of the National Science Foundation operated under cooperative agreement by 
Associated Universities, Inc.  Karl G.\ Jansky Very Large Array (VLA)} in L, C, and 
X band (Table~\ref{t:ic10vla}). Observations were performed in the continuum mode (by now
obsolete, after the upgrade to the WIDAR correlator) using a standard correlator setup 
with two IFs of 50\,MHz bandwidth each, centered at $1385.100$ and $1464.900$\,MHz, with 
all four correlations ($RR$, $LL$, $RL$, $LR$) recorded. We used J0059+581 as a phase 
calibrator to determine antenna gain solutions for amplitudes and phases, as well as 
the instrumental polarization leakages. As the primary calibrators, we used J0134+329 
(3C~48) and J0518+165 (3C~138) to prescribe the flux scale and the absolute position 
angle of the polarization E-vectors.

We followed standard data reduction procedures using {\small AIPS}. The observations 
in all bands and configurations were first edited and calibrated
separately and then self-calibrated in phase. Next, the data from all the 
observational sessions for a particular frequency band were combined and self-calibrated again.
In the final stage, imaging with  the {\small AIPS} task {\small IMAGR} was performed for all
the Stokes parameters using Briggs's robust weighting of 1 and $-5$,  where ``robust=5'' 
($-5$) is equivalent to natural (uniform) weighting. The obtained Stokes Q and U maps 
were combined to form maps of polarized intensity,  again corrected for the positive 
bias resulting from the Ricean noise distribution, and polarization angle. The obtained 
rms noise levels for total and polarized intensities are given in Table \ref{t:ic10vla}.

As IC\,10 has a relatively large angular optical size ($\approx$7$\arcmin$),
interferometric observations may lead to missing large-scale structure and
underestimation of the galaxy's flux (cf.\ Sect.~\ref{s:intro}). In order to
eliminate this effect, the data from VLA and Effelsberg in C and X bands were merged. 
The merging was performed within the Miriad package (Sault et~al. \citeyear{sault95},
task {\small IMMERGE}). The radio maps obtained in this way for I, Q and U
Stokes parameters, PI, and polarization angle were used in the further 
analysis (Sect.~\ref{s:results}).

\begin{figure*}
\centering
\includegraphics[clip,width=0.49\textwidth]{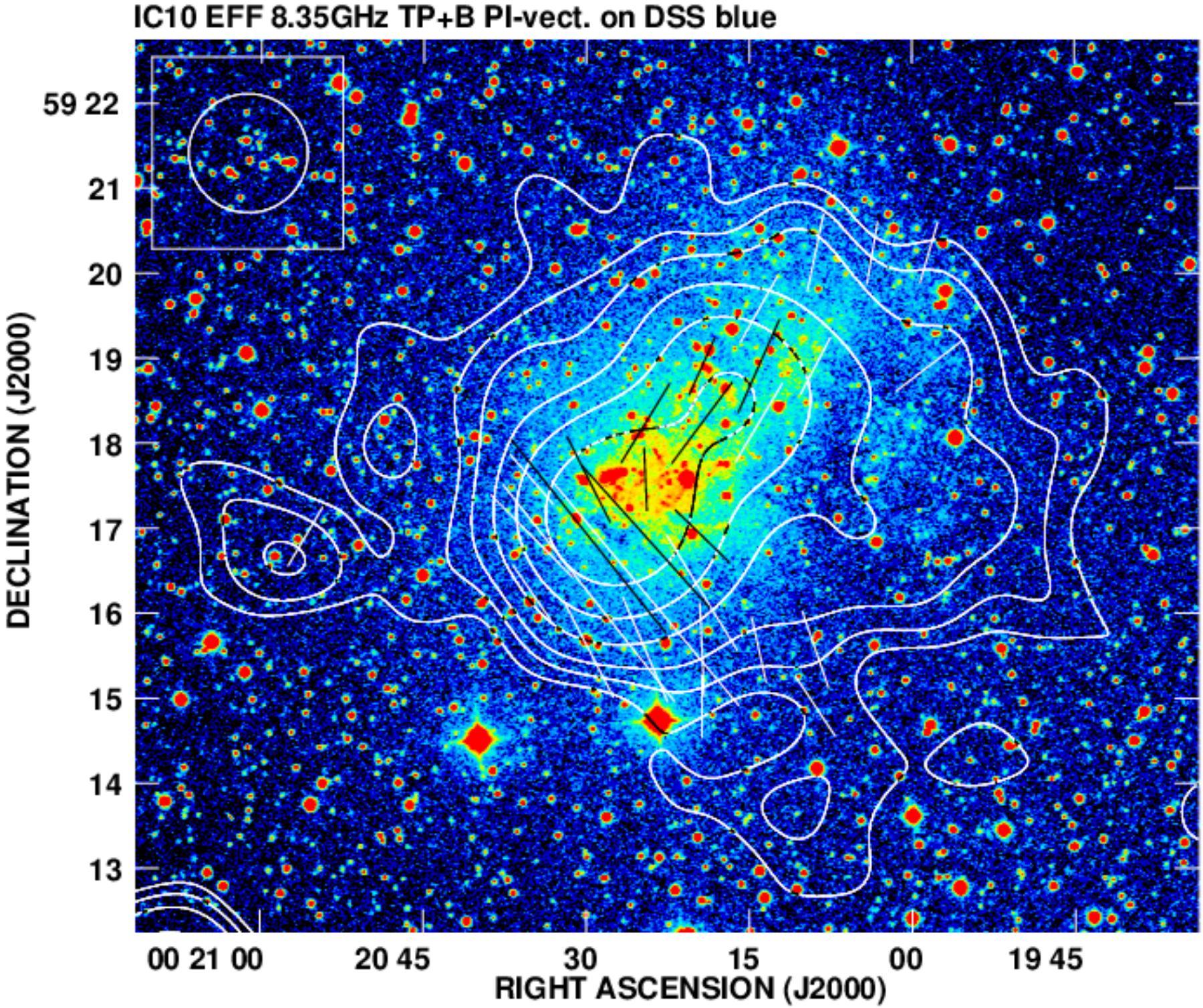}
\includegraphics[clip,width=0.49\textwidth]{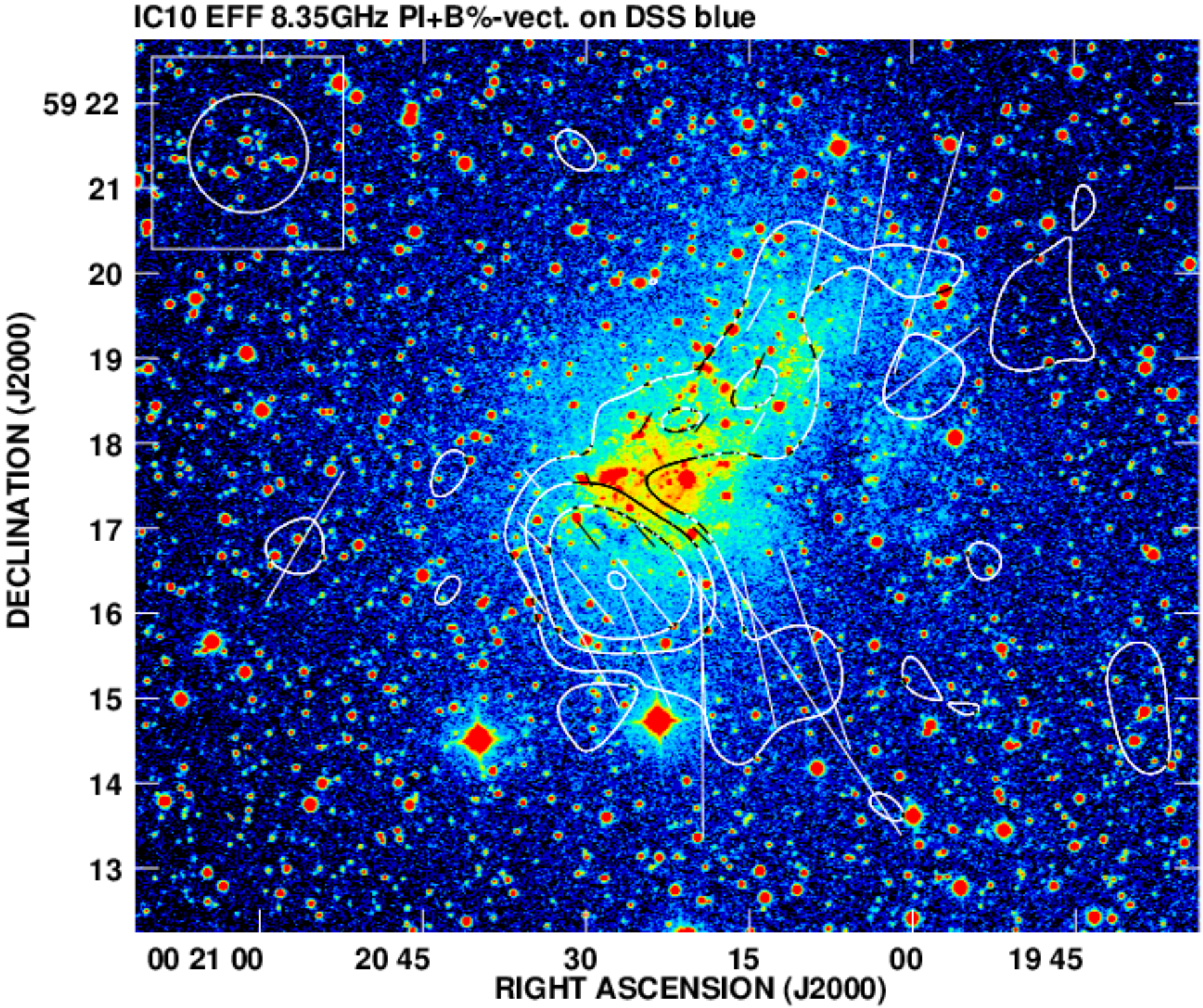}
\caption{Left: total-power contours and B-vectors of polarized intensity of IC\,10 at $8.35$\,GHz
superimposed on a DSS blue image. The contour levels are ($-3$, 3, 5, 8, 16, 32, 64) $\times$
500\,$\mu$Jy\,beam$^{-1}$. A vector of $10\arcsec$ length corresponds to a polarized intensity of 50\,$\mu$Jy\,beam$^{-1}$.
Right: contours of polarized intensity and B-vectors of polarization degree of IC\,10 at $8.35$\,GHz
superimposed on a DSS blue image. The contour levels are (3, 5, 8, 16) $\times$ 72\,$\mu$Jy\,beam$^{-1}$.
A vector of $10\arcsec$ length corresponds to a polarization degree of $0.7$\%. The resolution
of the maps is $84\arcsec\times 84\arcsec$ HPBW.}
\label{f:i10_3cm}
\end{figure*}

\section{Results}
\label{s:results}
\subsection{Low-resolution data}
In Figure \ref{f:i10_6cm} we present the total-power map of IC\,10 from the Effelsberg observations
at $4.85$\,GHz. The bright radio emission is morphologically oriented along the optical main body
of the galaxy, the maximum corresponding to the giant \ion{H}{2} region in the south, well visible in
the image in blue optical color. The low-level emission (at 3--5 times the noise level) 
extends far away from the galaxy's main stellar body in each direction. The three farthest 
elongations of the radio emission are background sources at 
R.A.(J2000)=$00^\mathrm{h}19^\mathrm{m}15\fs 0$, decl.(J2000)=$59\degr 19\arcmin 00\arcsec$; 
R.A.=00$^\mathrm{h}21^\mathrm{m}0\fs 8$, decl.=59$\degr$16$\arcmin$59$\arcsec$,
and R.A.=$00^\mathrm{h}20^\mathrm{m}8\fs 7$, decl.=59$\degr$13$\arcmin$27$\arcsec$),
well visible in the high-resolution NVSS map and our VLA map at $1.43$\,GHz (see below).
Another background source, located farther away and detached from the disk, is located at
R.A.=$00^\mathrm{h}21^\mathrm{m}10\fs 0$, decl.=$59\degr 11\arcmin 30\arcsec$.

The PI at $4.86$\,GHz is distributed into several major regions (Fig.~\ref{f:i10_6cm},
right panel).  The first of them is located in the southern part of the brightest \ion{H}{2} 
region. Emission at this place was also observed by Chy\.zy et~al.\ (\citeyear{chyzy03}) at 
$10.45$\,GHz. The orientations of B-vectors in both data sets are similar in this region and 
perpendicular to the galaxy's optical major axis (with $PA=125\degr$, but notice that the disk 
definition is uncertain for IC\,10; see Sect.~\ref{s:intro}). However, in the $4.85$\,GHz data 
an additional component (No.~2 in Fig.~\ref{f:i10_6cm}) with  a ``radial'' orientation
of B-vectors can also be seen. Another region of PI covers the northwest portion of the
optical disk and shows B-vectors roughly aligned with it. A hint of such a
magnetic field was found by Heesen et~al.\ (\citeyear{heesen11}) in their C-band JVLA observations 
of IC\,10. The other two PI regions (3 and 4) are located at the outskirts of the radio total 
emission up to about 1\,kpc distance from the optical edge of the disk. The B-vectors in these 
regions have a coherent pattern, with mostly radial orientations relative to the major axis 
of the optical disk. The characteristic valley in the PI visible in the east 
part of IC\,10, between the disk and the polarized component (3), is likely due to the effect of 
beam depolarization caused by perpendicular orientations of the magnetic fields in the disk 
and galactic outskirts. It  turns out that this galaxy, despite its small linear size, 
possesses a halo of weak, diffuse polarized emission, not detected in previous studies 
(Chy\.zy et~al.\ \citeyear{chyzy03}; Heesen et~al.\ \citeyear{heesen11}).

The total-power flux at $4.85$ GHz integrated over the whole radio extent of the galaxy is 
$222 \pm 9$\,mJy (Table \ref{t:ic10eff}), leaving out the four background sources. 
Heesen et~al.\ (\citeyear{heesen11}) found a flux density of 145\,mJy (extrapolated to 
$4.85$~GHz) from their interferometric data, which are not sensitive to
radio emission on large angular scales, which is 34\% deficient compared with our
Effelsberg maps. The polarized flux is $6.3\pm 0.4$\,mJy, which results in a polarization
degree of $2.8\% \pm 0.2$\,\%.

In Fig.\,\ref{f:i10_3cm} a higher-resolution (84$\arcsec$) radio map of IC\,10 from the Effelsberg data at
$8.35$\,GHz is shown. Similarly to $4.85$\,GHz, the maximum of emission again coincides with the brightest
\ion{H}{2} complex in the galaxy and the strong emission is found along the galaxy disk. Furthermore,
a characteristic diffuse emission to the west of the optical disk is visible, covering a long chain
(up to $3\arcmin$ in length) of H$\alpha$-emitting filaments. The map confirms similar features found at
$10.45$\,GHz by Chy\.zy et~al.\ (\citeyear{chyzy03}).

In the PI map (Fig.~\ref{f:i10_3cm}, right) there is a clear maximum of emission
corresponding to region 1 in the $4.85$\,GHz map (Fig.~\ref{f:i10_6cm}). The peak has no
bright H$\alpha$ counterpart, but coincides with the nonthermal superbubble of $48\arcsec$ size, found
there at R.A.=$00^\mathrm{h}20^\mathrm{m}28\fs 0$, decl.=$59\degr 16\arcmin 48\arcsec$ by Yang \& Skillman
(\citeyear{yang93}). The orientation of the B-vectors in this region is similar to those observed at $4.85$\,GHz,
which suggests no strong Faraday rotation effects there (see Sect.~\ref{ss:IC10RM}). There are
some patches of polarized emission along the optical disk at $8.35$\,GHz; however, no extraplanar (diffuse)
polarized signal is observed, in contrast to the $4.85$\,GHz detection in regions 3 and 4.

The integration of total radio emission gives a total-power flux at $8.35$\,GHz of
$183 \pm 8$\,mJy and a polarized flux of $2.5\pm 0.5$\,mJy. The average degree of 
polarization is $1.4$\%, slightly lower than at $4.85$\,GHz, which is probably due to
less extended emission observed at $8.35$\,GHz (e.g.\ no detection of
polarized emission in regions 3 and 4 that are seen at $4.85$\,GHz).

\subsection{High-resolution Data}
\label{ss:IC10HighResolutionResults}
\subsubsection{Synchrotron Envelope at 1.43\,GHz}

\begin{figure*}
\centering
\includegraphics[clip,width=0.49\textwidth]{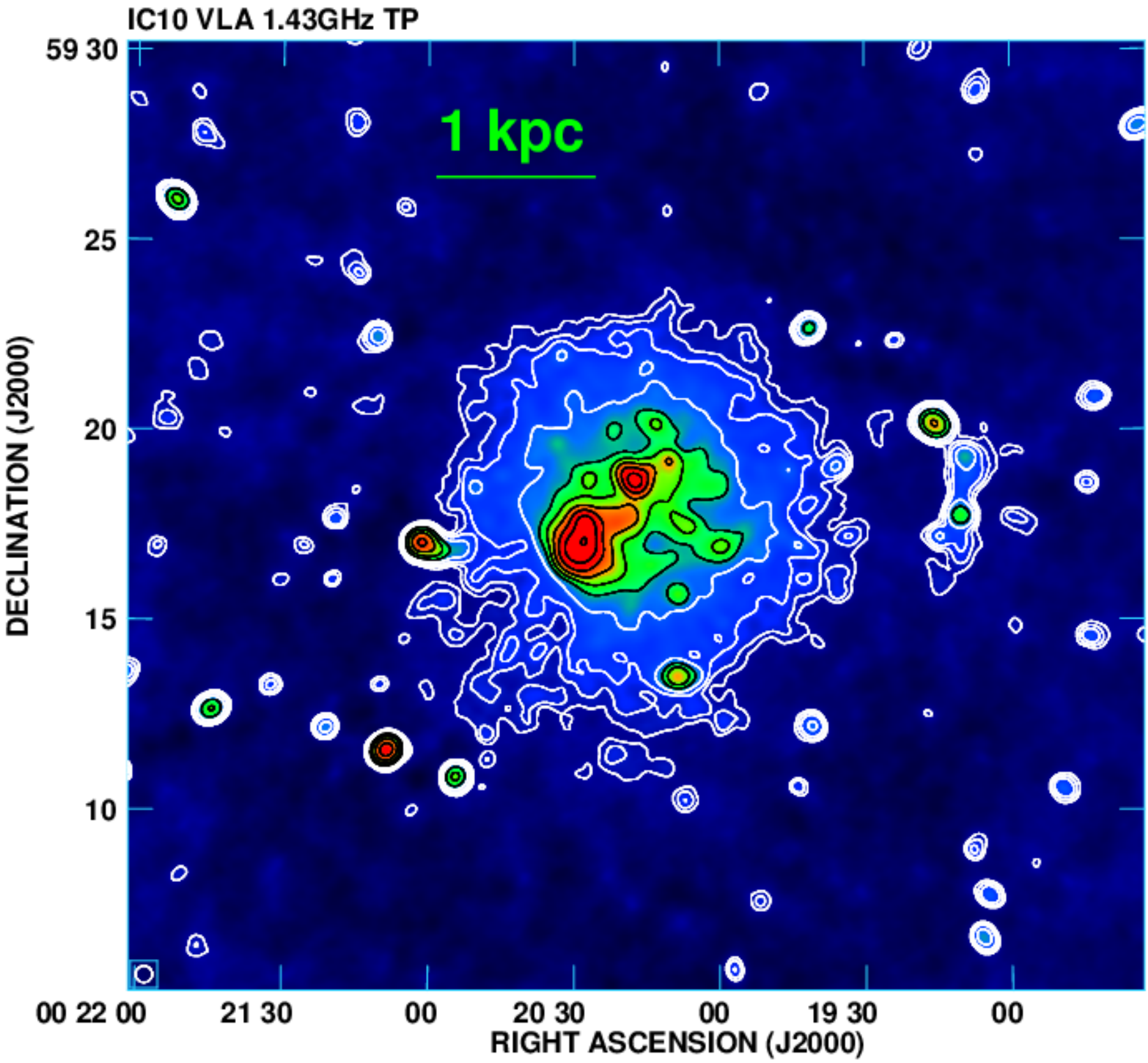}
\includegraphics[clip,width=0.49\textwidth]{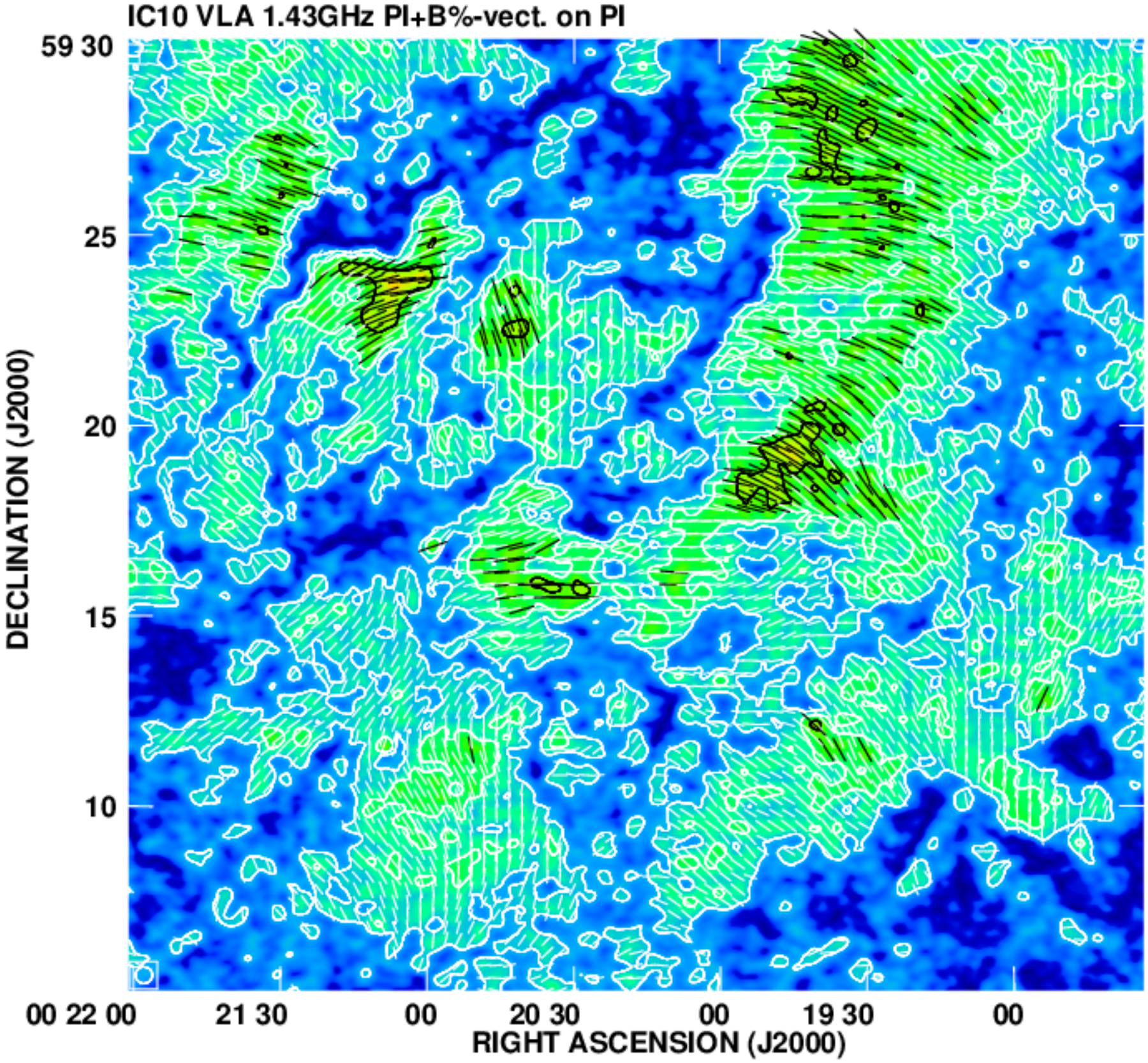}
\caption{
Left: total-power contours of IC\,10 at $1.43$\,GHz superimposed on the total-power image.
The contour levels are (3, 5, 8, 16, 32, 64, 128, 256, 512, 1024) $\times$ 28\,$\mu$Jy\,beam$^{-1}$.
Right: contours and B-vectors of polarized intensity of IC\,10 at $1.43$\,GHz superimposed on
the polarized intensity map. The contour levels are (3, 5, 8) $\times$ 15$\mu$Jy\,beam$^{-1}$. A vector
of $10\arcsec$ length corresponds to a polarized intensity of 25\,$\mu$Jy\,beam$^{-1}$.
The resolution of the $1.43$\,GHz radio maps is $26\arcsec\times 26\arcsec$ HPBW.
}
\label{f:i10vlaL26}
\end{figure*}
\begin{figure*}
\centering
\includegraphics[clip,angle=0,width=0.47\textwidth]{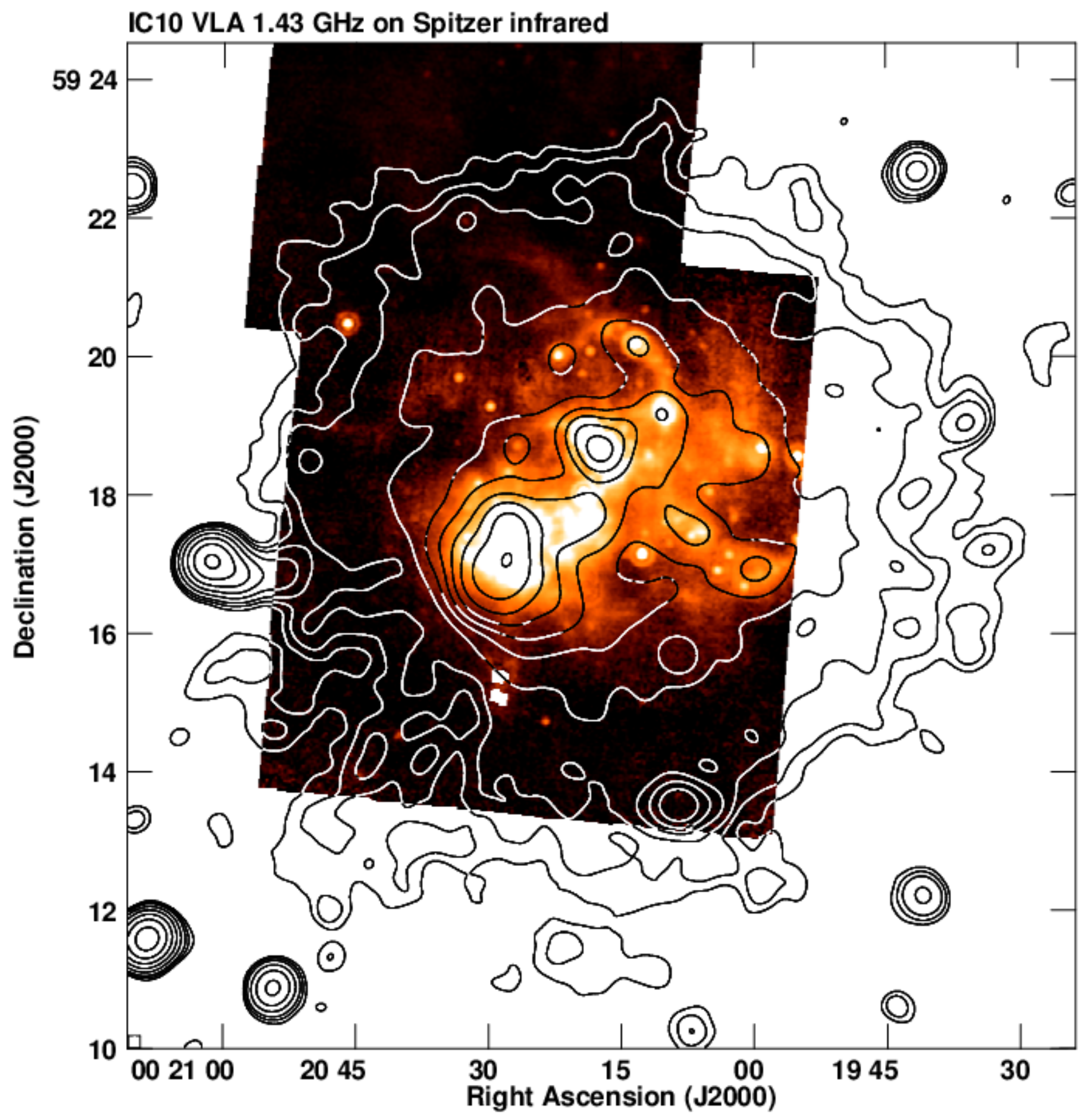}
\includegraphics[clip,angle=0,width=0.52\textwidth]{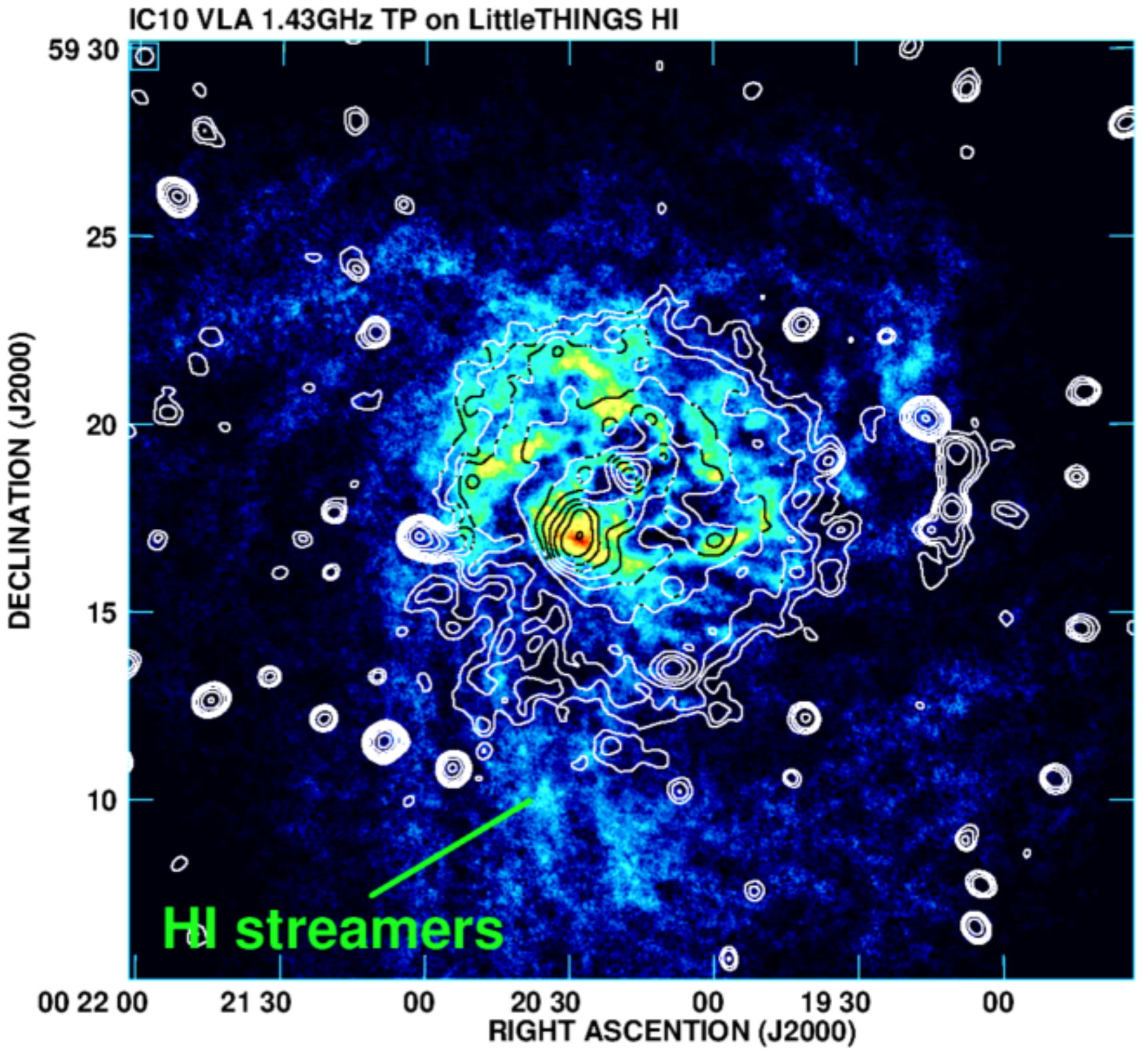}
\caption{
Left: total-power contours of IC\,10 at $1.43$\,GHz (see Fig. \ref{f:i10vlaL26}, 
left panel) superimposed on the {\em Spitzer} map at $24\,\mu$m (Bendo et al. 
\citeyear{bendo12}). The {\em Spitzer} map resolution is $5\farcs 3\times 5\farcs 3$ HPBW.
Right: total-power contours of IC\,10 at $1.43$\,GHz (see Fig. \ref{f:i10vlaL26}, left panel) 
superimposed on the \ion{H}{1} map (Hunter et~al.\ \citeyear{hunter12}). The \ion{H}{1} map 
resolution is $8\farcs 44\times 7\farcs 45$ HPBW.
}
\label{f:vlaTP_HI}
\end{figure*}

\begin{figure*}
\centering
\includegraphics[clip,width=0.49\textwidth]{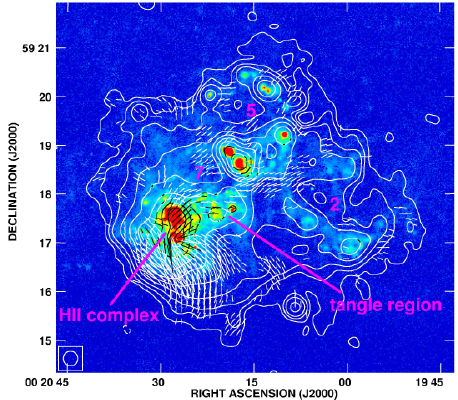}
\includegraphics[clip,width=0.49\textwidth]{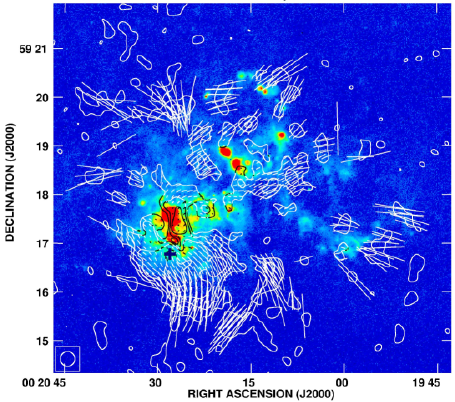}
\caption{Left: total-power contours and B-vectors of polarized intensity of IC\,10 (from the combined VLA
and Effelsberg data) at $4.86$\,GHz superimposed on an H$\alpha$ image (Gil de Paz et~al.\ \citeyear{gildepaz03}).
The contour levels are ($-3$, 3, 5, 8, 16, 32, 64, 128, 256, 512) $\times$ 17\,$\mu$Jy\,beam$^{-1}$. 
A vector of $10\arcsec$ length corresponds to a polarized intensity of about 33\,$\mu$Jy\,beam$^{-1}$.
The map resolution is $18\farcs 0\times 18\farcs 0$ HPBW. \ion{H}{1} holes 2, 5, and 7 are denoted (see
Wilcots \& Miller \citeyear{wilcots98}). Right: contours of polarized intensity and B-vectors 
of polarization degree of IC\,10 (from combined VLA and Effelsberg data) at $4.86$\,GHz 
superimposed on an H$\alpha$ image (Gil de Paz et~al.\ \citeyear{gildepaz03}).
The contour levels are (3, 5, 8, 16) $\times$ $6.5$\,$\mu$Jy\,beam$^{-1}$. A vector of $10\arcsec$ 
length corresponds to a polarization degree of $6.25$\%. The map resolution is $18\farcs 0\times 18\farcs 0$ HPBW.
The synchrotron superbubble is marked by a cross. The linear size of the superbubble is $48\arcsec$.
}
\label{f:i10_6_tpHalpha}
\end{figure*}

The sensitive low-frequency VLA observations of IC\,10 at $1.43$\,GHz reveal a very extended
total-power radio emission of the object (Fig.~\ref{f:i10vlaL26}, left). The envelope extends to
about $1.4$\,kpc ($6\arcmin$) in radius, much further than the stellar disk visible in Ks-band 
(Fig. \ref{f:i10_6cm}) and all the radio continuum emission at various frequencies published to date. 
We do not find matching H$\alpha$-emiting gas, but this can suffer from strong foreground extinction (the 
galactic latitude of IC\,10 is about $-3\fdg 3$). Because also the comparison of infrared emission 
observed by {\em Spitzer} at $24\,\mu$m does not reveal the galaxy envelope to the extent visible 
at 1.43 GHz (actually, the infrared-emitting disk is roughly two times smaller; Figure \ref{f:vlaTP_HI}, 
left panel), the envelope must be of nonthermal origin (see also Sect. \ref{s:thermal}).

We note that the galaxy in the $1.43$\,GHz map is surrounded by a weak ``negative bowl,'' 
indicating that missing zero spacings may prevent the detection of some very extended emission 
even at $1.43$\,GHz. Therefore, we suspect that the radio envelope of IC\,10 could be actually even larger. 
Much lower frequency observations are needed, e.g., with LOFAR to determine the full extent 
of the synchrotron halo in IC\,10. Low frequencies are preferred as they can reveal synchrotron emission generated by 
an old population of CR electrons spiraling along weak magnetic fields far away from star-forming regions.

The synchrotron envelope of IC\,10 is quite symmetric and smooth. While it extends 
farther than the distributions of main bulk of young and old stars, traced by {\it Spitzer }
$24\,\mu$m and the Ks-band data, its size in the east-west direction is similar to that of the 
neutral hydrogen (Fig.~\ref{f:i10vlaL26}). Farther away from the disk, where a network of 
\ion{H}{1} streamers and extensions can be seen, there are no traces of radio signal. 
The \ion{H}{1} data reveal many complex structures that result from some counterrotating 
components within the disk (Wilcots \& Miller \citeyear{wilcots98}) and remnants of past 
gravitational interaction (Ashley at al.\ \citeyear{ashley14}; Nidever et al.\ \citeyear{nidever13}). 
The synchrotron emission seems not to be associated with the \ion{H}{1} features and has 
its own more symmetrical morphology.

The total-power flux of IC\,10 measured at $1.43$\,GHz is $377\pm 11$\,mJy.

\subsubsection{Milky Way Contribution in Polarized Signal at 1.43\,GHz}

The distribution of the PI of IC\,10 at $1.43$\,GHz is another surprise of
this galaxy (Fig.~\ref{f:i10vlaL26}, right). There is a lot of polarized signal visible in the map,
but it is not directly related to IC\,10. In fact, the entire primary beam of the telescope is filled with
patches of diffuse polarized signal at a level of at least 45\,$\mu$Jy\,beam$^{-1}$. This, combined with the 
galactic latitude of IC\,10, which is about $-3\fdg 3$, leads one to suspect that the observed PI 
possibly comes from the Milky Way. It has been difficult to assess on the basis of the 
available data whether there is actually any polarized emission at $1.43$\,GHz coming from IC\,10. New
wideband VLA observations, e.g.\ in L band (1--2\,GHz), and application of  the rotation measure (RM) synthesis
technique, which for this frequency range would result in a Faraday depth resolution of about 40\,rad m$^{-2}$,
may allow us to separate the Milky Way foreground contribution from IC\,10.

These findings are quite peculiar ones, constituting the strongest manifestation of Milky Way
foreground emission in observations of any external galaxy  yet. A hint of additional polarized signal
not directly related to the investigated galaxy was found in NGC\,7331, which is located significantly
farther away from the Galactic plane ($-20\fdg 7$) than IC\,10 (Heald et al. \citeyear{heald09}).
Detection of strong polarized emission from the Milky Way portends difficulties in obtaining
and interpreting polarized signals from galaxies observed at extremely low radio frequencies,
now starting to be covered by the LOFAR instrument (Mulcahy et al.\ \citeyear{mulcahy14}).

\subsubsection{High-frequency Data and X-shaped Halo Field}

\begin{figure*}
  \begin{minipage}[c]{0.76\textwidth}
    \includegraphics[width=\textwidth,clip]{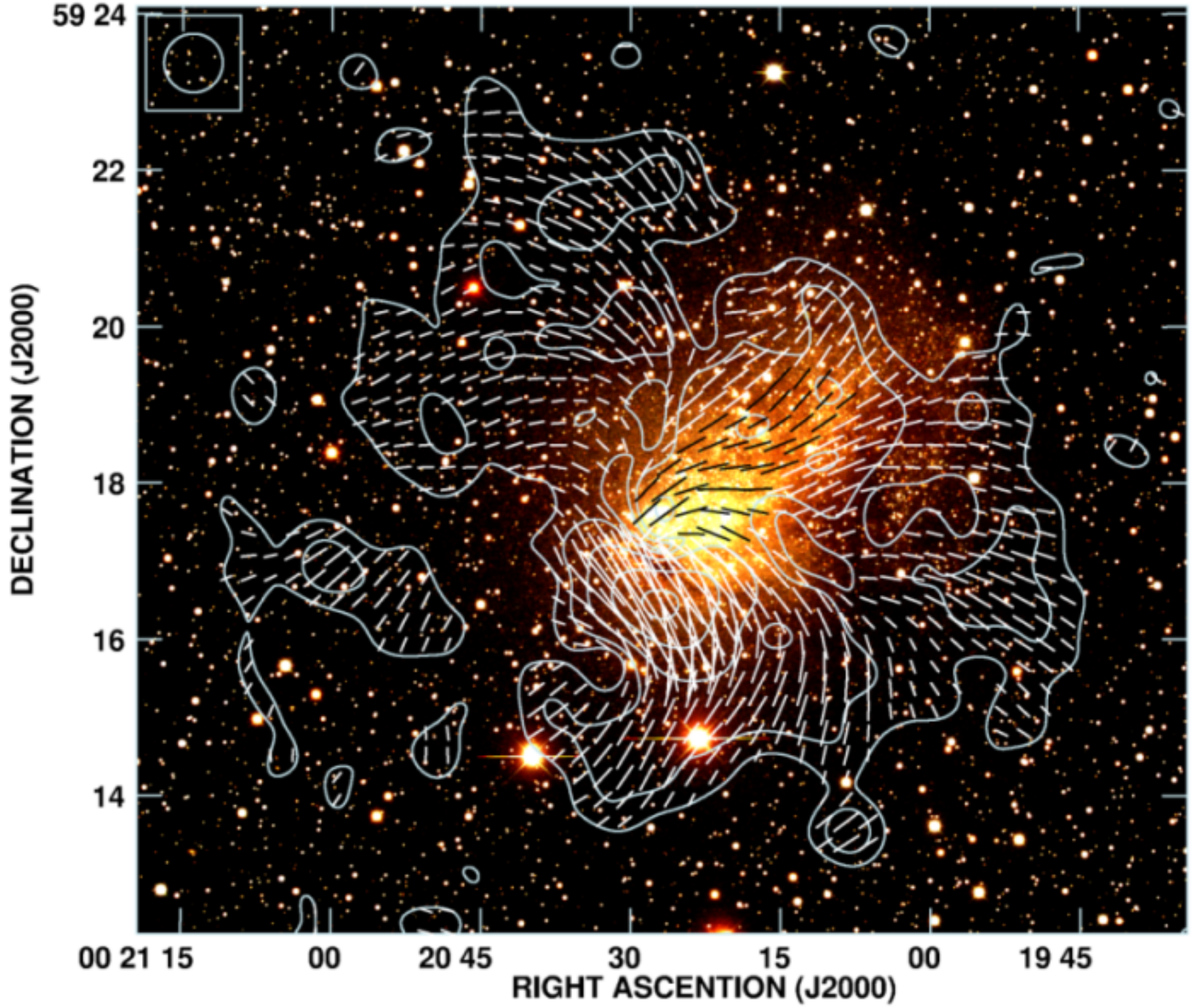}
  \end{minipage}\hfill
  \begin{minipage}[c]{0.23\textwidth}
    \caption{
Contours and B-vectors of polarized intensity (from combined VLA and
Effelsberg data) of IC\,10 convolved to the resolution of 45$\arcsec$
at $4.86$\,GHz and superimposed on the true-color optical image (from P.\  Massey/Lowell
Observatory and K.\ Olsen/NOAO/AURA/NSF). The contour levels are 
(3, 6, 12, 24, 40) $\times$ 11\,$\mu$Jy\,beam$^{-1}$. A vector of $10\arcsec$ length
corresponds to a polarized intensity of about 38\,$\mu$Jy\,beam$^{-1}$.
}
\label{f:i10_6_pi45}
  \end{minipage}
\end{figure*}

The total-power high-resolution ($18\arcsec$) map obtained from the combined VLA and 
Effelsberg data at $4.86$\,GHz is presented in Fig.~\ref{f:i10_6_tpHalpha} (left), 
superimposed on an H$\alpha$ image. Many radio intensity maxima correspond well to the bright
\ion{H}{2} regions. A more diffuse-like emission fills the entire optical disk
and extends farther to the northeast and southwest. Only the southeastern edge 
of the disk does not show such extension; - here the radio emission quickly drops to 
the noise level. This is surprising as this region is very close to the giant \ion{H}{2} 
complex where one would expect an efficient production of CR electrons. Something must 
prevent diffusion of CRs into this direction.

In Fig.~\ref{f:i10_6_tpHalpha} the PI map from the combined VLA and
Effelsberg data at $4.86$\,GHz is presented. The most prominent polarized signal found in the
object is located to the south and southwest of the giant \ion{H}{2} complex.
While the orientations of the B-vectors are perpendicular to the galactic major
axis in this region, they subsequently change to radial orientations in the
area farther to the south. The same pattern could be discerned in the Heesen 
et~al.\ (\citeyear{heesen11}) data, but here the polarization is also visible in other 
regions. Beyond the southern part of the galaxy, the distribution of the PI 
is patchy and apparently avoids any star-forming sites. \ion{H}{2} regions
actively form massive stars and produce enhanced turbulent motions that can effectively 
destroy ordered magnetic fields. We find  a similar process at work in the spiral arms 
of massive galaxies (e.g.\ Chy\.zy et~al.\ \citeyear{chyzy07b}).

The analysis of  the degree of polarization confirms this finding. The highest degrees 
of polarization, reaching values of up to 50\%, are found in regions of low star 
formation activity, i.e.\ in the outskirts of the galaxy. In the  nonthermal superbubble, 
we find values of $\approx$25\%. The lowest degree of polarization
of about 0.5\% is coincident with the giant \ion{H}{2} complex.
The polarized flux measured for the whole galaxy at $4.86$\,GHz is 4.9 $\pm$
0.6\,mJy, which results in a mean polarization degree of 2.6\% $\pm$ 0.3\,\%.

In order to obtain a clearer global picture of the distribution of the PI 
in IC\,10, we convolved the PI map  with a Gaussian kernel to a lower 
angular resolution of 45$\arcsec$ (Fig.~\ref{f:i10_6_pi45}). Astonishingly, 
a pattern of global magnetic fields emerges in this dwarf irregular galaxy, 
extending far beyond the optical disk into the halo! In its farthest extensions, 
found  northeast and southwest of the main stellar body, the orientations of 
the B-vectors are radial, suggesting the presence of a halo magnetic field.
The ordered magnetic field pointing away from the center of the galaxy
displays a partly distorted ``X-shaped'' morphology. Note that this structure 
is also seen in Fig.~\ref{f:i10_6cm}. There is also a quite distinct magnetic 
field component aligned with the main optical body of the galaxy ($PA=125\degr$).
X-shaped halo magnetic fields have so far been found only in late-type spiral 
edge-on galaxies (e.g.\ Soida et~al.\ \citeyear{soida11}), but they have never 
been detected in any dwarf irregular galaxy. The magnetic structure of
IC\,10 partly resembles the X-pattern in the late-type spiral galaxy NGC\,4631, which
possesses a large radio halo with radial field lines advected in a strong galactic wind
(Mora \& Krause \citeyear{mora13}).  See Sect.~\ref{s:bfieldstructure} for
further discussion.

\begin{figure*}
\centering
\includegraphics[clip,width=0.49\textwidth]{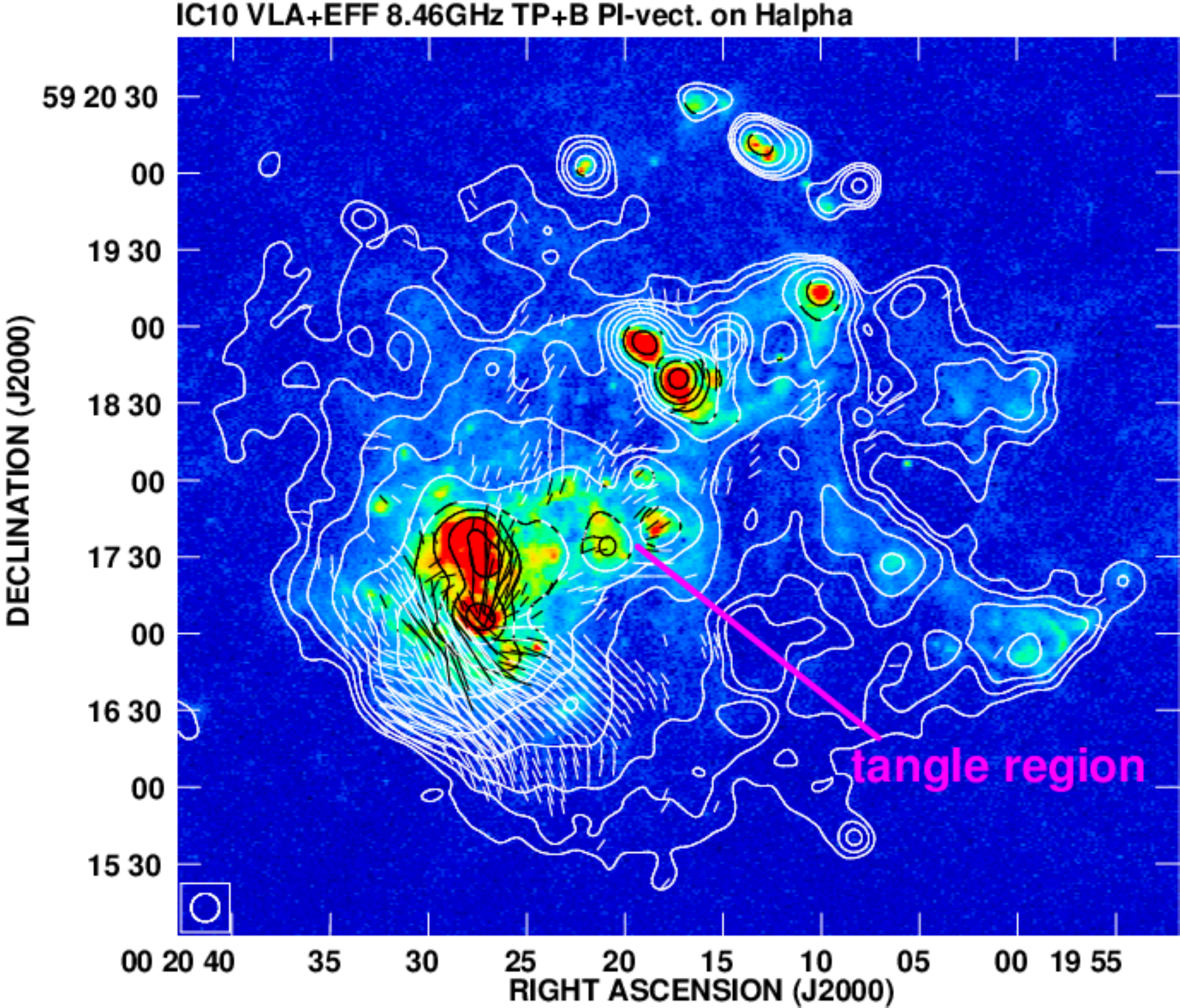}
\includegraphics[clip,width=0.49\textwidth]{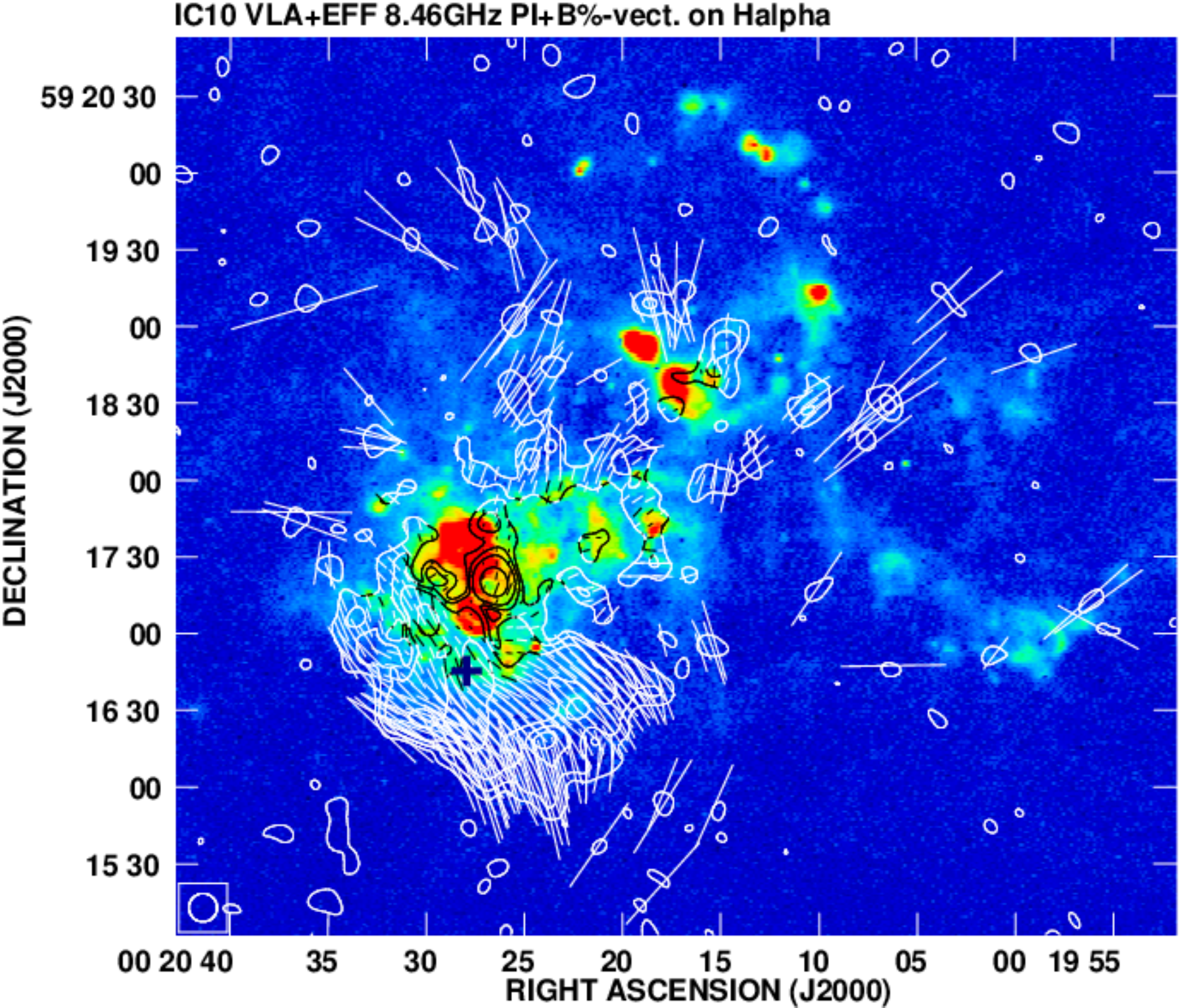}
\caption{
Left: total-power contours and B-vectors of polarized intensity of IC\,10
(from combined VLA and Effelsberg data) at $8.46$\,GHz superimposed on the
H$\alpha$ image (Gil de Paz et~al.\ \citeyear{gildepaz03}). The contour
levels are ($-3$, 3, 5, 8, 16, 32, 64, 128, 256, 512, 1024) $\times$
10\,$\mu$Jy\,beam$^{-1}$. A vector of $10\arcsec$ length corresponds to a polarized
intensity of about 25\,$\mu$Jy\,beam$^{-1}$ (not corrected for the primary beam
attenuation). The map resolution is $11\farcs 0\times 11\farcs 0$ HPBW.
Right: contours of polarized intensity and B-vectors of polarization
degree of IC\,10 (from combined VLA and Effelsberg data) at $8.46$\,GHz
superimposed on the H$\alpha$ image (Gil de Paz et~al.\ \citeyear{gildepaz03}).
The contour levels are (3, 5, 8, 16) $\times$ $3.5$\,$\mu$Jy\,beam$^{-1}$.
A vector of $10\arcsec$ length corresponds to the polarization degree of about
$7.14$\%. The map resolution is 11$\arcsec$ HPBW.
}
\label{f:i10_mrg_3cm}
\end{figure*}

In Figure \ref{f:i10_mrg_3cm}, we present the total-power radio continuum
map of IC\,10 at $8.46$\,GHz, which has the highest linear resolution (44\,pc) of
all the maps included in this work.  The corresponding map of the linear
polarization (Fig.~\ref{f:i10_mrg_3cm}) is the most sensitive image
(3.5\,$\mu$Jy\,beam$^{-1}$) obtained for this object to date.  Both the total and
PI distributions are very similar to those at 4.86\,GHz 
(compare Fig.~\ref{f:i10_6_tpHalpha}). As for the 4.86\,GHz data, there is 
an apparent anticorrelation between the PI and the bright 
\ion{H}{2} complexes in the disk, whereas the opposite is observed for the 
total intensity. We find the same situation in the so-called ``tangle region'' 
of Thurow \& Wilcots (\citeyear{thurow05}), which is composed of many distinct,
compact \ion{H}{2} regions immersed in diffuse H$\alpha$ emission. Weak polarized 
emission is detectable at the edges of this region and reveals magnetic field 
vectors directed outward from this area (Fig.~\ref{f:i10_mrg_3cm}).

The bulk of the polarized emission is found in the region of the nonthermal superbubble
and the area adjacent to the southwest.  Here the field lines are probably  unrelated
to the superbubble as they cross the bubble from northeast to southwest and continue farther
to the south. The B-vector orientations  mostly agree with those at 4.86\,GHz, 
indicating that Faraday rotation is small and we observe the intrinsic magnetic field orientation.

From the high-resolution data at $8.46$\,GHz we find integrated flux densities of $154\pm 4$\,mJy and
$4.1\pm 0.3$\,mJy, for total-power and polarization, respectively. Hence, the
mean degree of polarization is $2.7\% \pm 0.2$\,\%. The spatially resolved
degree of polarization has its minimum with 2--3\% close to the giant
\ion{H}{2} complex, whereas the maximum values are found close to the southern edge of
the halo with values of $\approx$30\%. Such  a high degree of polarization could be 
caused by either compression or stretching  of magnetic field lines. Both
processes may be active in this area, whereas they are suppressed in the nearby 
giant \ion{H}{2} complex. Gravitational interaction or gas accretion  can cause 
gaseous streamers and bulk flows of \ion{H}{1} gas,  leading to compression and
stretching of the magnetic field lines (see Sect.~\ref{s:bfieldstructure}).

\subsection{Distribution of Spectral Index and Nonthermal Emission}
\label{s:thermal}

\begin{figure}
\centering
\includegraphics[clip,width=0.48\textwidth]{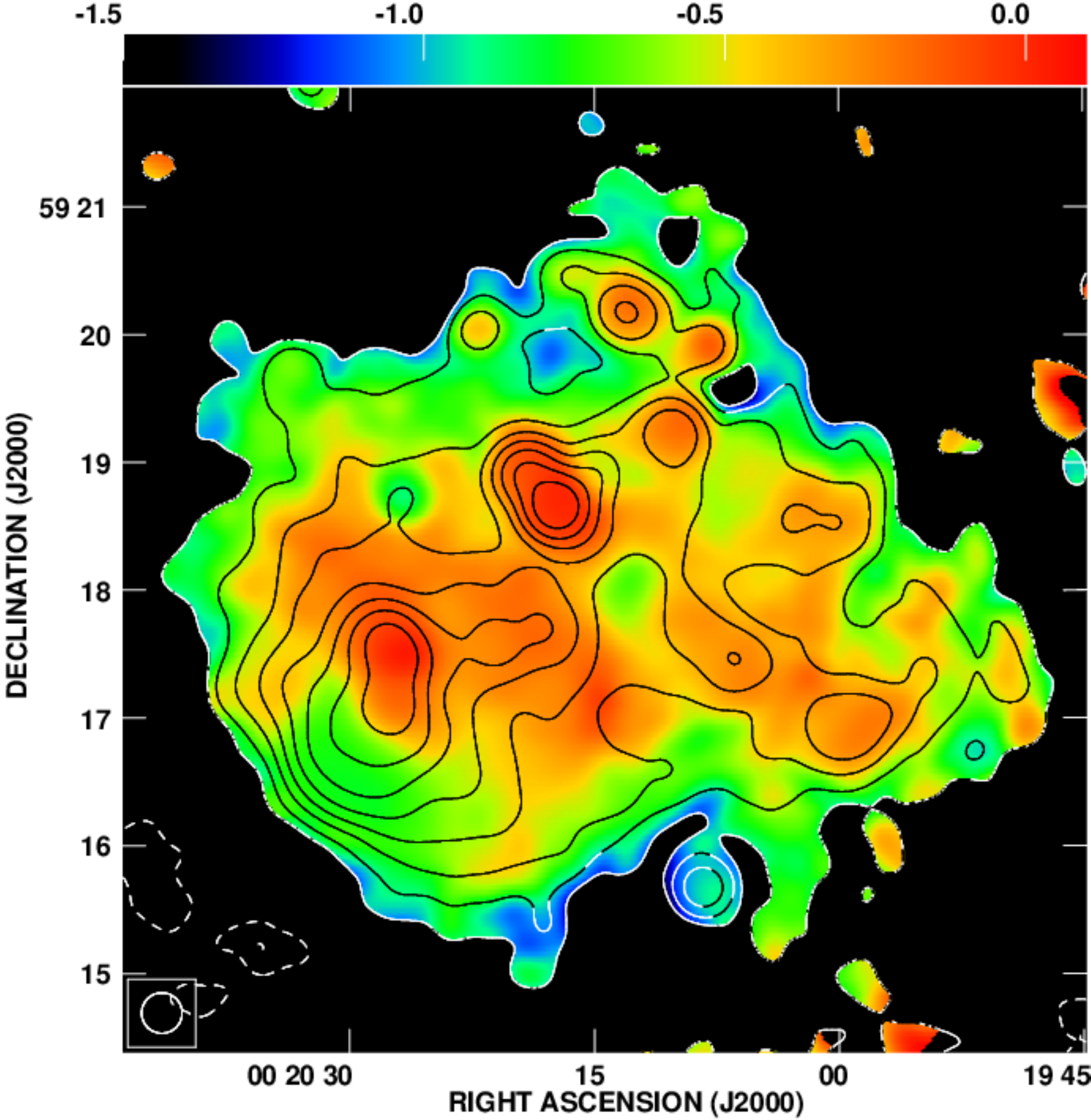}
\caption{Radio spectral index distribution between $1.43$ (Robust 0) and 
$4.86$\,GHz (merged VLA and Effelsberg data) in IC\,10 (colors). Both the 
maps of total intensity were convolved to a common beam of 19$\arcsec$. The contours represent
the total-power map at $4.86$\,GHz; the levels are ($-5$, $-3$, 5, 8, 16, 
32, 64, 128, 256, 512) $\times$ 17$\,\mu$Jy\,beam$^{-1}$.}
\label{f:i10_SPIX}
\end{figure}

In Fig.~\ref{f:i10_SPIX} we present the spectral index distribution between 
$1.43$\,GHz (imaged with robust 0) and $4.86$\,GHz (from the combined VLA
and Effelsberg data). Both maps were first convolved to the common resolution
of $19\arcsec$ and  clipped at intensities below 5 times the rms noise level.
Flat spectral indices at $-0.1$ to $-0.2$ are cospatial with
bright, compact \ion{H}{2} regions,  which suggests that they are dominated by
thermal emission. Indeed, IC\,10 is known to manifest a significant thermal fraction
of radio emission (about 50\% at $10.45$\,GHz; Chy\.zy et~al.\ \citeyear{chyzy03}).
Similar flat spectra were also observed by Heesen et~al. 
(\citeyear{heesen11}) in their high-resolution spectral index map obtained with
the multifrequency synthesis method. However, their map is limited to just the galaxy's
brightest part. In the spectral index map presented in this paper a more diffuse 
component of radio emission, located mainly in between star-forming regions, 
can also be discerned. Such regions show steeper spectral indices,  ranging 
between $-0.4$ and $-0.5$, indicating a mix of nonthermal (synchrotron) and 
thermal (free--free) emission.

Toward the edges of the galaxy, the spectral index steepens to a value of 
about $-1$, which means that these regions are dominated by synchrotron emission.
The same  applies to the region of the nonthermal superbubble, where the
spectral index has a value of about $-0.8$.

The spectral index map shows a northeast-southwest asymmetry across the optical disk
with flatter spectral indices found on the southwestern side. Such a pronounced 
asymmetry is not visible in the $1.43$\,GHz total intensity map, indicating
that the spectral asymmetry is due to strong thermal emission from regions of
recent star formation on the southwestern side of the disk. Such emission 
comes particularly from the ``tangle region'' and from the long curved H$\alpha$ 
extension (around R.A.=00$^\mathrm{h}20^\mathrm{m}00^\mathrm{s}$, 
decl.=59$\degr$17$\arcmin$50$\arcsec$) filled with young shells and gas filaments.

\begin{figure}[t]
\centering
\includegraphics[clip,width=0.48\textwidth]{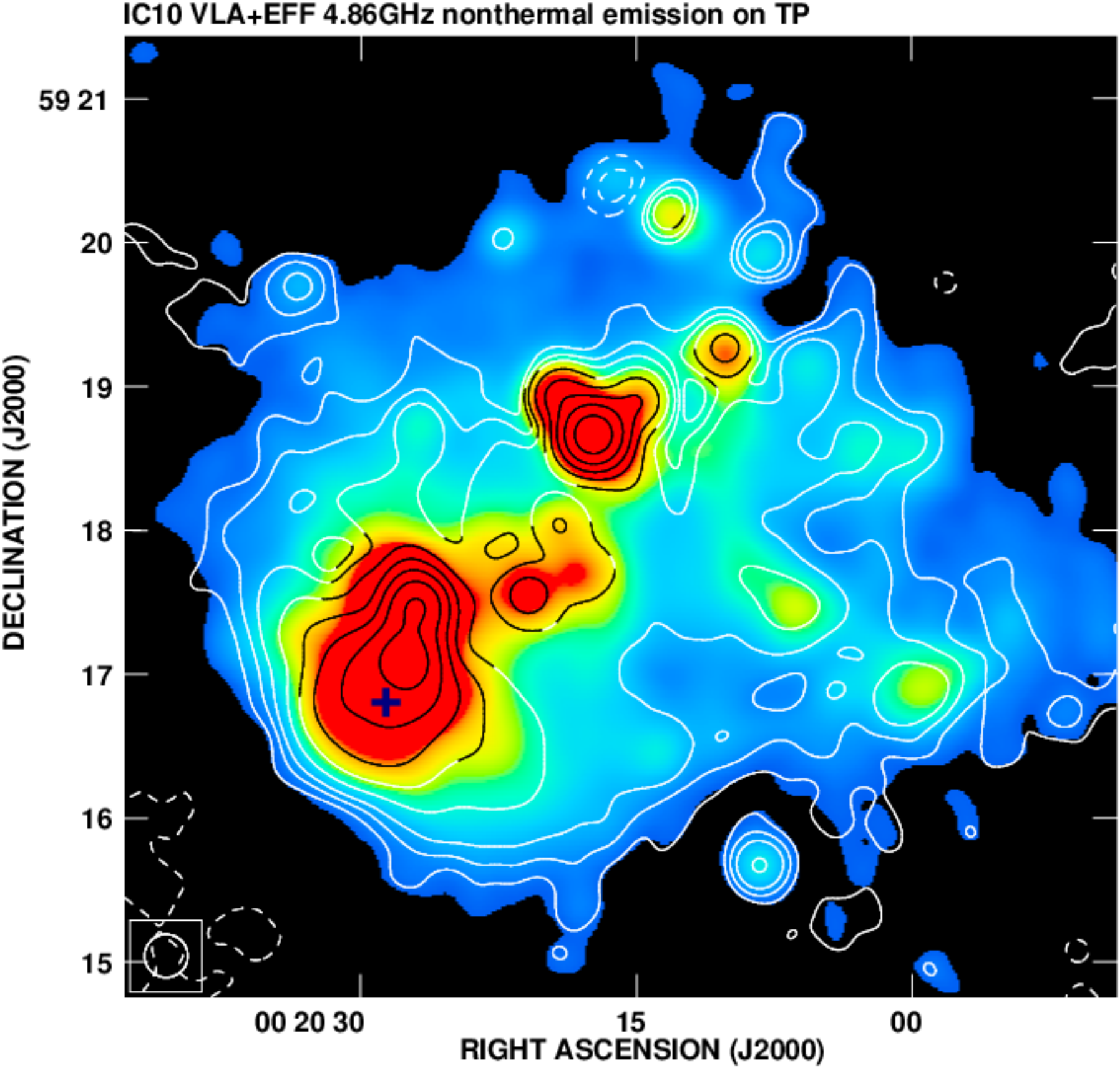}
\caption{Distribution of nonthermal radio emission in contours superimposed 
on the total-power radio emission at $4.86$ in IC\,10. The contours levels 
are: ($-5$, $-3$, 3, 5, 8, 16, 32, 64, 128, 256) $\times 25\,\mu$Jy\,beam$^{-1}$.
The map resolution is 18$\arcsec$.
}
\label{f:i10_NTH}
\end{figure}

A separation of the thermal and nonthermal components of the radio emission is essential 
for the determination of the magnetic field strength (Sect.~\ref{ss:IC10BfieldStrength}) 
and for the analysis of the relation between the magnetic field and other ISM phases. 
In our approach we use the estimation of the mean H$\alpha$ dust extinction  
from our previous paper of Chy\.zy et~al.\ (\citeyear{chyzy03}) to correct our H$\alpha$ map 
for dust attenuation. The theory of free-free emission is then used (Caplan \& 
Deharveng \citeyear{caplan86}) to derive from the corrected H$\alpha$ map the predicted 
thermal radio flux map of IC\,10 at 4.86\,GHz. This map is then subtracted from the 
total radio map at 4.86\,GHz to obtain the distribution of nonthermal emission (Fig.~\ref{f:i10_NTH}). 

The nonthermal emission of IC\,10 fills the entire optical body of the galaxy, as 
found in star-forming spiral galaxies (e.g.\ Chy\.zy et~al.\ \citeyear{chyzy07a}). 
It is particularly strong in the region of the nonthermal superbubble (as expected)
and in the whole southern polarized region. However, it is also strong in
the southern and northern \ion{H}{2} complexes, hinting at a connection
between star formation, production of CRs, and amplification of magnetic fields.
Here the spectral index is flatter than in the polarized region. In the western 
area, there is a significant extension of nonthermal emission, which follows the 
chain of \ion{H}{2} regions. In contrast, the nonthermal emission in 
the opposite (northeastern) direction is weaker and extends only half as far out into 
the halo. In this  area, there are also fewer \ion{H}{2} regions, 
and hence also less sources able to produce CRs and synchrotron emission. This is 
clearly confirmed by the steeper spectral index on the northeastern side of the 
disk (Fig.~\ref{f:i10_SPIX}). These findings are also supported by X-ray observations, 
- which show confined $10^6$\,K gas in the southern part and the main disk of 
IC\,10 (Wang et al.\ \citeyear{wang05}). The hot gas has probably
already escaped from the western chain of (older) \ion{H}{2} regions, as well as from the
eastern galaxy part. In the southern part the population of W-R and other massive
stars is still able to maintain the hot gas and balance the cooling of X-rays.

\begin{figure}[t]
\centering
\includegraphics[clip,width=0.48\textwidth]{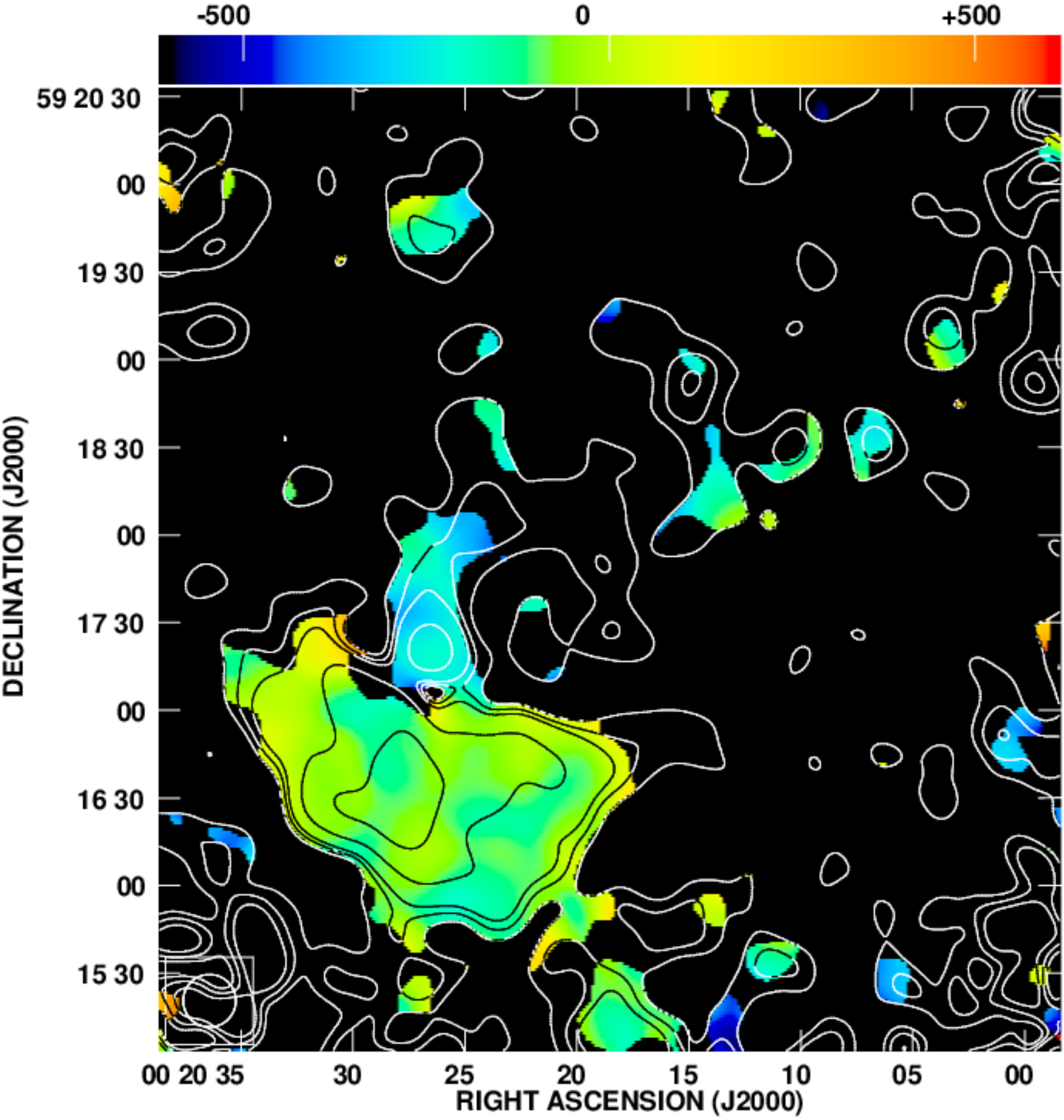}
\caption{
Distribution of Faraday RM in IC\,10 (in rad\,m$^{-2}$, not 
corrected for the foreground RM) from the data at $4.86$ and $8.46$\,GHz in 
IC\,10 (colors). The contours represent the polarized intensity map at $8.46$\,GHz; the levels
are: (4, 6, 8, 16, 32) $\times$ 8\,$\mu$Jy\,beam$^{-1}$. The map resolution is 18$\arcsec$.
}
\label{f:i10_RM}
\end{figure}

In contrast to the $4.86$\,GHz emission, we find that the $1.43$\,GHz emission,
which is dominated by the nonthermal component, has a much higher degree
of symmetry, forming an almost spherical halo. It would seem that  an older population 
of CR electrons fill the radio envelope of IC\,10 in a much more uniform way,
requiring efficient CR transport. As will be shown in Sect. \ref{s:winds}, 
the convection timescale for the CRs to be transported out to the edge of the 
halo at 1.5 kpc distance is 25\,Myr (assuming a wind speed of about 60\,km\,s$^{-1}$, 
see Sect. \ref{s:winds}). The origin of the oldest CR electrons in the halo thus 
predates the current starburst and suggests that they are stemming from an 
earlier epoch of star formation (see Sect. \ref{ss:IC10BfieldStrength}).

\subsection{Distribution of RM}
\label{ss:IC10RM}

In the present study, for the first time, we obtained a map of the RM distribution
in IC\,10 (Fig.~\ref{f:i10_RM}). The map was calculated using the observations at $4.86$ 
and $8.46$\,GHz convolved to  a common angular resolution of 30$\arcsec$. Both 
polarization maps were clipped at intensities below 4 times the rms noise level. 
The resulting typical uncertainty of RM is about 68\,rad\,m$^{-2}$ for a 
signal-to-noise ratio of (S/N) of 4 and smaller for larger S/N. The most characteristic 
RM structure is a large region in the south of the galaxy corresponding to the 
nonthermal superbubble and the polarized extension to the southwest from it. In this area
the RM is approximately $-86\pm 26$\,rad\,m$^{-2}$,  with some
variation over the whole region. In the southern giant \ion{H}{2} complex it 
reaches values around $-250\pm 33$\,rad\,m$^{-2}$. The RM distribution is more 
patchy in other regions. Large values of RM in the most southern location, 
at declinations below 16$\degr$, are due to  an increased rms noise in the proximity 
of the primary beam edge at $8.46$\,GHz.
 
Nonvanishing absolute RM values, averaged in regions larger than the telescope beam,
might suggest the existence of regular\footnote{Meaning magnetic field vectors
with the same direction and sense} magnetic fields in IC\,10. This would shed 
a new light on the generation and evolution of the magnetic field in this object and 
challenge the current understanding of the magnetic field generation mechanism in dwarf 
galaxies (Chy\.zy et~al.\ \citeyear{chyzy03}). However, IC\,10 is located almost 
in the Milky Way's plane, which may contribute to the total RM signal, acting 
as a foreground Faraday screen. From the all-sky RM map of 
Oppermann et al.\ (\citeyear{oppermann12}) we found that for the Galactic 
coordinates  of IC\,10 the interpolated Galactic Faraday depth (foreground RM) 
is about $-98\pm 61$\,rad\,m$^{-2}$. Large uncertainty results from the low 
resolution of the map (about $0.5\degr$) and variation of the Faraday depth close
to the Galaxy plane. As this value is close to the one determined from our RM map
in the direction of the superbubble, the Galactic foreground can almost fully 
explain the obtained RM signal, and we do not have any conclusive evidence for a regular
magnetic field in IC\,10.

After subtracting the foreground RM from the value determined for the giant 
\ion{H}{2} complex, we obtain the residual value of about $-150\pm 69$\,rad\,m$^{-2}$.
This may arise from some local enhancement of the Milky Way foreground signal or 
might be intrinsic from within IC\,10. Systematic negative values
of RM indicate regular magnetic fields oriented mostly away from the observer.
With a density of thermal electrons of $0.1$\,cm$^{-3}$ and the synchrotron path length 
of 1\,kpc, such an RM value would indicate regular fields in this region of $\approx 2\pm1\,\mu$G.
Taking the path length two times smaller results in two times larger regular  field strength.

With the present data we do not see any evidence for a galaxy-scale regular magnetic 
field in IC\,10. We only found hints of local and relatively weak regular fields 
of about $2\,\mu$G strength in the giant \ion{H}{2} complex. The existence of
coherent structures of RM over large regions in IC\,10 can further be investigated by independent 
sensitive spectropolarimetric observations, preferably at low radio frequencies to probe 
the extensive radio envelope. Application of RM synthesis is necessary to 
enable the separation of polarized signal coming from the Milky Way and IC\,10. Such 
observations are feasible with LOFAR.

\subsection{Relation of Radio Emission of IC\,10 to Other Galaxies}
\label{ss:radiofir}

\begin{figure}
\centering
\includegraphics[angle=0,width=0.48\textwidth,clip]{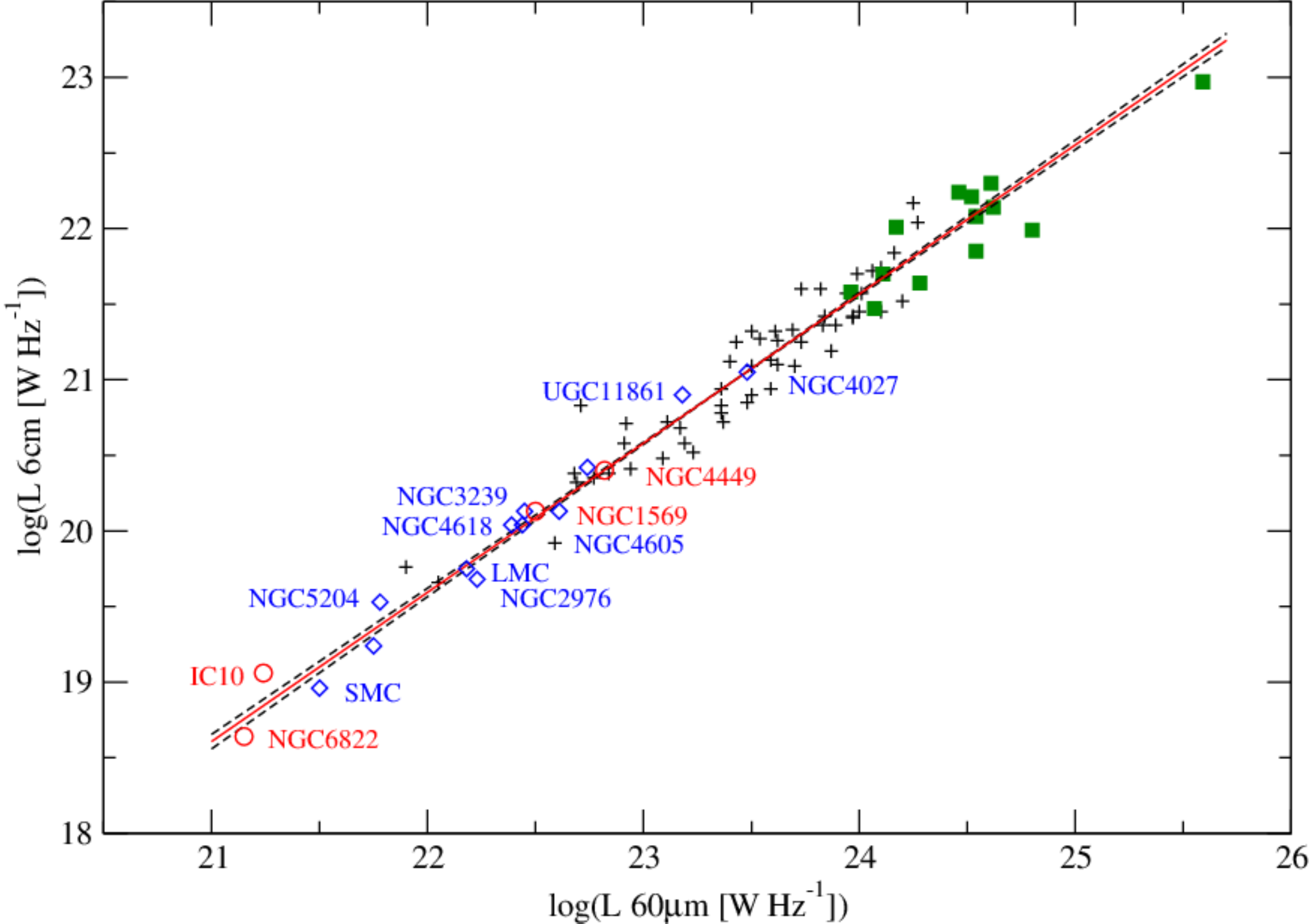}
\caption{Radio--far-infrared correlation diagram for dwarf galaxies (circles), 
Magellanic-type and peculiar galaxies (diamonds), interacting objects (squares), 
and bright spiral galaxies (crosses). The luminosities at $4.85$\,GHz ($6.2$\,cm) and 
at $60\,\mu$m are used, respectively. The solid line is a bisector fit to all 
(83) galaxies, while the dashed lines represent simple X vs.\ Y and Y vs.\ X regressions. 
Uncertainties in measurements are of the order of the symbols' sizes. The index 
of the power-law from the bisector fit is $0.99\pm 0.02$.}
\label{f:radiofir}
\end{figure}

We compare the radio emission of IC\,10 with other galaxies using the
radio--far-infrared (RFIR) diagram based on luminosities at $4.85$\,GHz and
$60\,\mu$m, respectively. For the infrared emission of IC\,10 we adopt a
value of $33.5$\,Jy from Fullmer \& Lonsdale (\citeyear{fulmer89}).
To construct the RFIR relation, we use dwarf galaxies (NGC\,6822,
NGC\,4449, NGC\,1569), low-mass objects that we recently observed 
(NGC\,2976, NGC\,3239, NGC\,4027, NGC\,4605, NGC\,4618, NGC\,5204,
UGC\,11861), and five other Magellanic-type galaxies (Jurusik et al.\
\citeyear{jurusik14}), 54 bright spirals from Gioia at al.\ (\citeyear{gioia82}), and 
13 objects from our compilation of interacting galaxies 
(Drzazga et al.\ \citeyear{drzazga11}).

Our investigation shows (Fig.~\ref{f:radiofir}) that IC\,10 does not deviate
from the power-law fit constructed for all 83 galaxies of various types.
It resides at the low-luminosity end of the relation, where galaxies of
low mass and small total SFR are located. In this respect it is similar to the SMC. 
However, IC\,10 has a high value of the surface
density of star formation and strong magnetic fields (Sect.~\ref{ss:IC10BfieldStrength}),
similar to the starbursting dwarfs NGC\,4449 and NGC\,1569, which are larger, 
more massive, and more luminous than IC\,10 (see Chy\.zy et al.\ \citeyear{chyzy11}).
The total radio and infrared power of IC\,10 is therefore not spectacular, but
the local properties of the ISM are probably much like those in more 
massive starbursting dwarfs. They all could be triggered by gravitational 
interactions with other objects (see Sect.~\ref{s:intro}).

\section{Discussion}
\label{s:discussion}

\subsection{Magnetic Field Strength}
\label{ss:IC10BfieldStrength}

Our observations revealed that IC\,10 possesses a diffuse, weak radio 
envelope and an unusual magnetic field morphology. In order to investigate
the properties of the magnetic field further, we calculate global and local values of its strength
and compare them  with other galaxies. To this end, we use the combined VLA and Effelsberg
$4.86$\,GHz data at $18\arcsec$ angular (equivalent to 70~pc) resolution. For the calculation of
the average magnetic field strength we assumed a thermal fraction of $0.5$ 
and a nonthermal spectral index of $\alpha_\mathrm{nth}=-0.6$, following Chy\.zy
et~al. (\citeyear{chyzy03}). To derive local values, we make use of the
nonthermal map (Fig.~\ref{f:i10_NTH}), assuming $\alpha_\mathrm{nth}=-0.6$ 
within \ion{H}{2} regions and $\alpha_\mathrm{nth}=-0.8$ outside of them (see Heesen 
et al. \citeyear{heesen15}). A disk thickness of 1\,kpc was used as the integration length
along the line of sight. The calculations were performed with the standard assumption 
of energy equipartition between the magnetic field and total CRs, assuming a 
proton-to-electron ratio of 100 (Beck \& Krause \citeyear{beck05}).

We find an average total magnetic field strength of B$_\mathrm{tot}=14\,\mu$G, which
is stronger than in other low-mass galaxies (see, e.g., Chy\.zy et~al.\ \citeyear{chyzy11}; 
Jurusik et~al.\ \citeyear{jurusik14}). With  the help of  the polarized emission, 
we also estimate the  strength of the ordered field as B$_\mathrm{ord}=2.3\pm 0.7$\,$\mu$G 
and that of the random field component as B$_\mathrm{ran}=13.3 \pm 4.0$\,$\mu$G. 
The degree of field order is $q=\mathrm{B_{ord}/B_{ran}}=0.17\pm 0.07$, which 
is lower than typically observed in spiral galaxies. Our value is, however, in 
close agreement with the mean degree of field order of $0.27\pm 0.09$,
estimated for a sample of interacting galaxies showing kinetically disordered disks
(Drzazga et~al.\ \citeyear{drzazga11}).

\begin{table}
\caption{Magnetic field strength and Degree of Field Order in IC\,10}
\begin{center}
\begin{tabular}{cccc}
\hline
\hline
Region     & B$_\mathrm{tot}$       & B$_\mathrm{ord}$     & Degree of  \\
           & $\mu$G                 & $\mu$G               & Field Order\\
\hline
Whole galaxy           & $13.5$ $\pm 4.1$ & $2.3 \pm 0.7$ & $0.17 \pm 0.07$  \\
Giant \ion{H}{2} complex      & $28.7\pm 8.6$ ($38.0^a$) & $1.6\pm 0.5$ & $0.06\pm 0.03$ \\
North \ion{H}{2} complex      & $28.4\pm 8.5$ ($39.9^a$) & $1.6\pm 0.5$ & $0.06\pm 0.03$ \\
Tangle region          & $15.4\pm 4.6$ & $1.3\pm 0.3$ & $0.08 \pm 0.05$ \\
\ion{H}{1} hole 2              & $9.0\pm 2.7$ & $1.0\pm 0.3$ & $0.11\pm 0.05$ \\
\ion{H}{1} hole 5              & $7.1\pm 2.1$ & $3.5\pm 1.1$ & $0.56\pm 0.24$ \\
\ion{H}{1} hole 7              & $10.4\pm 3.1$ & $2.7\pm 0.8$ & $0.27\pm 0.12$ \\
Nonth. superbubble     & $22.0\pm 6.6$ & $3.2\pm 1.0$ & $0.15\pm 0.06$ \\
South PI extension     & $13.0\pm 3.9$ & $4.7 \pm 1.4$ & $0.39\pm 0.17$ \\
Northern outskirts     &  $8.6 \pm 2.6$  & $2.6\pm 0.8$ & $0.32\pm 0.14$ \\
Southern outskirts     &  $10.5\pm 3.2$ & $2.3\pm 0.7$ & $0.22\pm 0.09$  \\
Halo edge ($1.4$\,GHz)   &   $7.0\pm 2.1$ & n.a.\ & n.a.\  \\
\hline
\end{tabular}
\end{center}
$^a$ - Local value of the magnetic field strength after subtracting the influence
of the background (disk) emission
\label{t:ic10Btot}
\end{table}

The strongest magnetic field is observed locally in the bright part of the southern 
giant \ion{H}{2} complex with B$_\mathrm{tot}=29\,\mu$G (Table \ref{t:ic10Btot}). 
A similar value is found in the northern \ion{H}{2} complex, which is also bright 
in H$\alpha$, but less extended. Generally speaking, regions with weak H$\alpha$ emission have
also weak field strengths, as, for instance, does the ``tangle region,'' where 
B$_\mathrm{tot}=15\,\mu$G (see Fig.~\ref{f:i10_6_tpHalpha} for a nomenclature of 
the regions). Inside holes of the \ion{H}{1} distribution, which are in the 
vicinity of the brightest star-forming regions (Wilcots \& Miller \citeyear{wilcots98}, 
see also Fig.~\ref{f:i10_6_tpHalpha}), the magnetic field strength is halved in 
comparison with the surrounding area. In the nonthermal superbubble, the magnetic field
is still strong (B$_\mathrm{tot}=22\,\mu$G) and  dominated by the random  component 
(the degree of field order $q=0.15$).\footnote{We refer to the superbubble as the 
entire structure along the line of sight (assumed here to be here 1~kpc) in the
direction of the superbubble. Heesen et al. \citeyear{heesen15} assumed
that the superbubble only extends 153~pc along the line of sight and thus
arrived at a higher magnetic field strength of $44\pm 8\,\mu$G.}
The superbubble is filled with CR electrons stemming from between one and a
few supernova explosions, which occurred approximately 1~Myr ago (see Heesen et
al.\ \citeyear{heesen15}). Because the nonthermal spectral index inside the
bubble is significantly steeper, compared with surrounding areas, it is
likely that the CR electrons are contained within and not able to freely escape.
In the adjacent area to the south of this \ion{H}{2} region, the magnetic
field  strength gradually decreases to B$_{\mathrm{tot}}=13\,\mu$G  while
becoming more ordered ($q=0.39$). If this region is also supplied with 
some CRs from the giant \ion{H}{2} complex, the rising field order may suggest 
stretching of magnetic fields and diffusion of CRs along them to the south. 
Because in the northern and southern outskirts of IC\,10 the magnetic fields have similar
properties (see Table \ref{t:ic10Btot}), the global outflow of magnetized plasma and stretching
of the magnetic field in the whole observed radio envelope can explain the observed field strengths.

We also determined the strength of the total magnetic field in areas
farthest away from the disk in the halo, which fall below the detection
threshold at $4.86$\,GHz. We used  the $1.43$\,GHz map and assumed that the 
thermal emission is negligible at the edge of the galaxy halo. For the synchrotron 
path length we adopted a value of $0.5$\,kpc, expecting a smaller path at the
halo edge. Under these conditions, our estimations give total magnetic field strengths in the
range of $6.8$--$7.1\,\mu$G, both in both the northern and southern halo, approximately 
$1.4$\,kpc away from the giant \ion{H}{2} complex. Adopting the less likely value of 1\,kpc 
for the path length results in diminishing the field strength by only 16\%.
In spite of the large distance from the galaxy disk, far  away from the warm
and hot ionized medium phases of the ISM, there is magnetized plasma with still strong 
magnetic fields.

While estimating the magnetic field strength  as described above, it was assumed 
that in IC\,10 synchrotron cooling dominates CR electrons' energy losses. But, in 
the galaxy's outskirts the inverse Compton scattering may become important and cool 
relativistic electrons, because the magnetic field energy density may become 
comparable to the photon energy density of the cosmic microwave background.
This is the case when the total magnetic field strength drops below
$3.25\,\mu {\rm G }(1+z)^2$. But according to our estimations in low-level 
radio emission regions of IC\,10 (to the north and to the south of the disk), using the
above assumptions, the magnetic field strength is 7--10\,$\mu$G (Table \ref{t:ic10Btot}).
This suggests that in the outer parts of the galaxy losses by inverse Compton scattering 
are always smaller than synchrotron losses, which justifies the way we estimate 
the radiative cooling of CR electrons. In the disk, inverse Compton losses are probably
higher owing to photons from the galaxy's radiation field, so that our estimate of 
the electron radiation losses in the disk should be treated as a lower limit.

\subsection{Magnetic Field Structure}
\label{s:bfieldstructure}

\begin{figure}
\centering
\includegraphics[clip,width=0.48\textwidth]{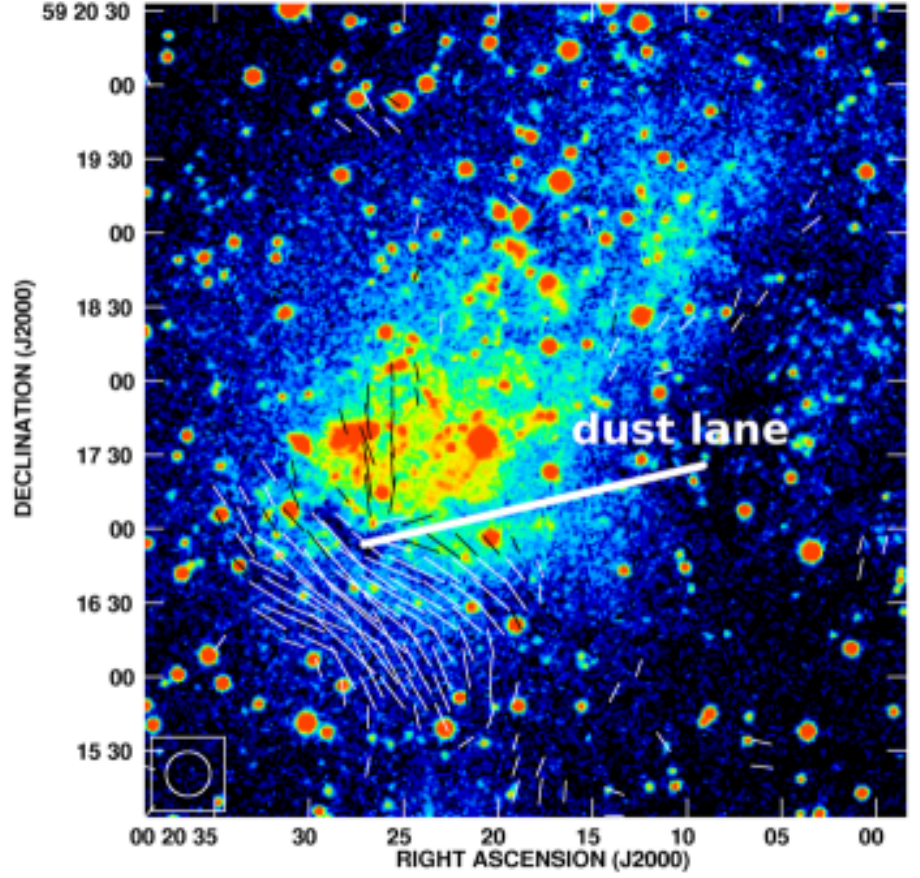}
\caption{Magnetic field vectors of polarized intensity (from the combined VLA and 
Effelsberg data of 18$\arcsec$ resolution) of IC\,10 superimposed on the DSS blue 
image. A vector of $10\arcsec$ length corresponds to the polarized intensity of 
about 38\,$\mu$Jy\,beam$^{-1}$. The position of the strong dust lane is indicated.}
\label{f:i10_BfieldDirections}
\end{figure}

As was shown in section \ref{s:results}, the magnetic topology in the outer disk
of IC\,10 resembles the X-shape --- a magnetic field structure dominated by the  radial
component, observed commonly in edge-on spiral galaxies (see, e.g.\ T\"ullmann et~al. 
\citeyear{tullmann00}; Soida et~al.\ \citeyear{soida11}). As the observed vectors at
$8.46$\,GHz (where the Faraday effects are small) have orientations in the radio halo
similar to those at $4.86$\,GHz, they likely represent the true orientations of the magnetic
field in projection onto the sky plane. In the northeastern part of the object, the B-vectors
apparently form an elongated and curved structure. A similar field curvature can be
seen in the opposite direction, around the polarized extension (Fig.~\ref{f:i10_6_pi45}).
There is also a weak trace of B-vectors roughly aligned with the galaxy optical disk
($PA=125\degr$). 

Numerical MHD simulations show that in spirals the observed X-structures 
of magnetic fields could arise from the quadrupolar mode of the large-scale 
($\alpha$--$\Omega$) dynamo (Beck et al.\ \citeyear{beck96}),
with the galactic wind transporting magnetic flux from the disk into the halo 
(Brandenburg et al.\ \citeyear{brandenburg93}; Moss et al.\ \citeyear{moss10}).
The poloidal component of the quadrupolar field alone cannot explain 
the X-shaped field, as according to the mean-field dynamo theory 
it is about 10 times weaker than the azimuthal disk field. Therefore, 
the vertical outflow is needed. Simulations of the CR-driven 
dynamo that include strong vertical winds (blowing with the speed exceeding 
100\,km\,s$^{-1}$) reveal the X-shaped structure in the edge-on view of spirals 
(Hanasz et al.\ \citeyear{hanasz09}).

For the first time we find a similar X-pattern of magnetic fields 
in the dwarf galaxy IC\,10, which is a stellar system with a highly irregular 
distribution of star-forming regions, does not reveal spiral
density waves, and has a highly disturbed \ion{H}{1} velocity field. Is the operation
of a large-scale dynamo still possible in IC\,10 to produce such fields?
The efficiency of generation of the magnetic field can be roughly estimated
 from the dynamo number, according to the following equation:
\begin{equation}
D\approx 9 \frac{h^{2}_{0}}{u^{2}_{0}} s \omega \frac{\partial \omega}{\partial s}
\end{equation}
where $h_0$ is the vertical scale height of the galactic disk, $u_0$ is the 
velocity of turbulent motions, $\omega$ is the angular velocity at the radial 
distance $s$ from the center of the galaxy, and $s \partial \omega / \partial s$ is the shear.
In our calculations we assume $s=0.5$\,kpc and a velocity of rotation of about 30\,km s$^{-1}$
at a distance of $0.8$\,kpc, where the rotation curve is flat (Wilcots \& Miller \citeyear{wilcots98}).
This gives a shear of 38\,km s$^{-1}$ kpc$^{-1}$. The velocity of turbulent
motions we approximate by the velocity dispersion, which is about 30--40\,km s$^{-1}$ in the
central part of the disk (Wilcots \& Miller (\citeyear{wilcots98}). These 
values yield a dynamo number $|D|\approx$4 which is below the critical value 
of $\approx$10. Therefore, the classical $\alpha$--$\Omega$ dynamo cannot
work in IC\,10.

The numerical MHD simulations by Siejkowski et al.\ (\citeyear{siejkowski12}) 
show that in a dwarf galaxy rotating with a velocity of about 30\,km\,s$^{-1}$ 
the CR-driven dynamo produces a regular magnetic field of only about $1\,\mu$G strength.
Owing to the slow rotation of the gas, the e-folding time for developing such a field
is quite long, about 800--1700\,Myr, and can hardly explain magnetic fields in IC\,10
as this is longer than the duration of the current starburst (10\,Myr). The 
magnetic structure seen in IC\,10 also does not resemble any numerical models 
simulated for dwarf galaxies (Dubois \& Teyssier \citeyear{dubois10}; Siejkowski 
et al.\ \citeyear{siejkowski12}; Siejkowski et al.\ \citeyear{siejkowski14}). None 
of the models predict magnetic fields of the strength and vertical alignment as we 
observe in IC\,10. It might be that gravitational interactions and gas accretion 
play an important role in the generation and evolution of the field in IC\,10, 
which was not the emphasis of numerical simulations of dwarf galaxies so far.

Therefore, we propose another possibility to explain the magnetic 
X-pattern in IC\,10: spreading and ordering of anisotropic magnetic fields 
by a global gas outflow, e.g.\ by a galactic wind (see Sect.~\ref{s:winds}). 
In this scenario, a large random component of the magnetic field is produced by
a small-scale fluctuating dynamo and locally ordered by gas flows. The
small-scale MHD dynamo can be sustained by the intensive phase of star
formation visible in H$\alpha$ emission. A comparison of the B-vectors with
the distribution of \ion{H}{2}-emitting gas  gives the impression that the 
whole magnetic structure could be dominated by outflows just from the southern 
giant \ion{H}{2} complex. This region is the youngest and most vivid 
star-forming area in the galaxy, as, e.g., it shows the bulk of W-R stars 
and harbors the bulk of the hot, X-ray-emitting gas (Wang et al.\ \citeyear{wang05}).
If this is the case, we might see in IC\,10 large fragments of magnetic
loops anchored to this region,  similarly to the ones found in the
30~Doradus giant star formation region in the Large Magellanic Cloud (Mao et al.\
\citeyear{mao12}). This would easily explain some curvatures of field lines
visible on different sides of the galaxy. However, this appealing idea cannot
easily explain the highly symmetric distribution of the radio intensity observed
at $1.43$\,GHz, which indicates that the CRs and the magnetic field originate
from the entire disk, rather than from only the southern part. One can speculate that 
the X-pattern is not directly related to the radio halo seen at $1.43$\,GHz, 
being a more recent phenomenon, associated mainly with the most prominent region of the
current star formation burst. The radio halo could  originate from the
outflow triggered by the previous star-forming activity dated to  have occurred
at least 150\,Myr ago (see Sect.~\ref{s:intro}).

The only polarized structure that appears to be unrelated to the global
X-shaped pattern is the polarized region in the southern part of the object. 
For a more precise analysis we corrected the magnetic field orientation at 
$4.86$\,GHz for Faraday rotation using a map of the RM between 
$4.86$ and $8.46$\,GHz. Owing to restricted polarized signal at $8.46$\,GHz, 
this was possible only in the southern part of the galaxy 
(Fig.~\ref{f:i10_BfieldDirections}). The orientations of the vectors closely 
resemble the low-resolution but high-frequency ($10.55$\,GHz) magnetic structure
obtained from the Effelsberg data (Chy\.zy et~al.\ \citeyear{chyzy03}). The magnetic 
vectors seem to be unrelated to either the H$\alpha$ or \ion{H}{1} emission but
are closely aligned with a long dust lane visible to the south
from the giant \ion{H}{2} complex (Fig.~\ref{f:i10_BfieldDirections}). Such
a configuration of magnetic field lines could  suppress the diffusion of CR 
electrons into areas southeast of the nonthermal superbubble (perpendicular 
to the field lines), which  would explain the strong gradient
of radio emission seen in this region at $1.43$\,GHz (Fig.~\ref{f:i10vlaL26}).

It is  also possible that some additional compression or stretching
forces are at work here  as well. According to recent investigations by 
Nidever et~al.\ (\citeyear{nidever13}), IC\,10 is
likely gravitationally interacting with another dwarf galaxy, which
could induce various gas flows modifying the magnetic field morphology.
Magnetic  field lines aligned with the most pronounced dust lane in the galaxy
might be  of a very recent origin, probably from the ongoing starburst, 
which started only 10\,Myr ago. Another process that could lead to field 
ordering is  ongoing accretion of primordial gas, proposed to explain 
the complicated \ion{H}{1} morphology of the object (Wilcots \& Miller 
\citeyear{wilcots98}).

Summarizing, there are probably several processes that shape the magnetic 
field structure in IC\,10: first, compression and stretching in the 
vicinity of the expanding nonthermal superbubble in conjunction with gas 
flows, either from accretion of primordial gas or induced by recent 
tidal interaction: second, gas outflows from compact star-forming
\ion{H}{2} regions with a range of intensities, leading to a very localized
(tens of parsec) velocity field; third, a galactic wind (1~kpc scale), driven
by stellar winds from young, massive stars and supernova explosions.
Spectropolarimetric observations in a wide range of frequencies are definitely 
needed to obtain more complete information on Faraday rotation and thus on intrinsic
orientations of the magnetic field and to draw more robust conclusions about
any external influences, possibly also responsible for the magnetic field structure in IC\,10.

\subsection{Magnetized Galactic Winds}
\label{s:winds}

\begin{figure}
\centering
\includegraphics[clip,width=.23\textwidth]{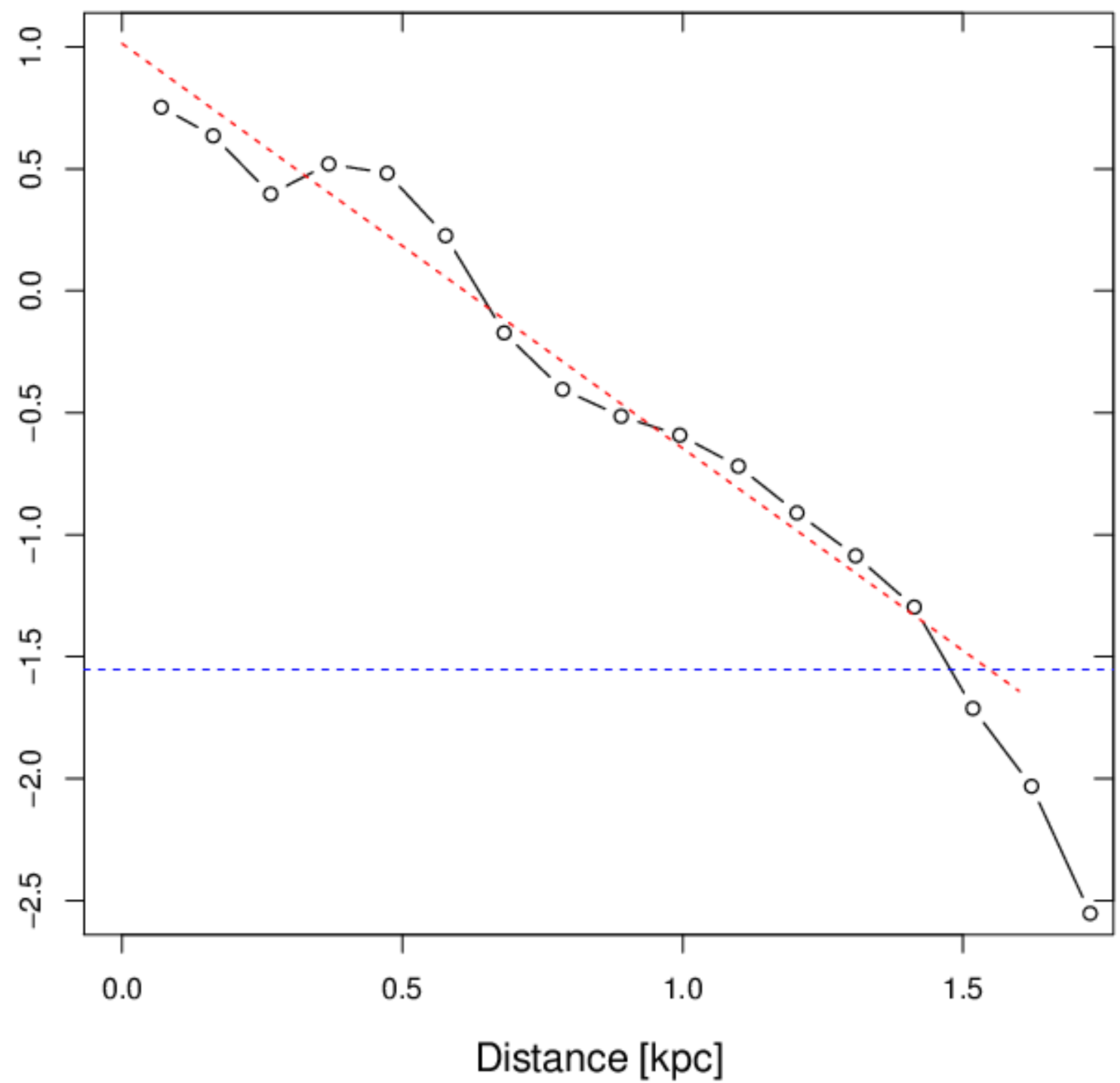} \includegraphics[clip,width=.23\textwidth]{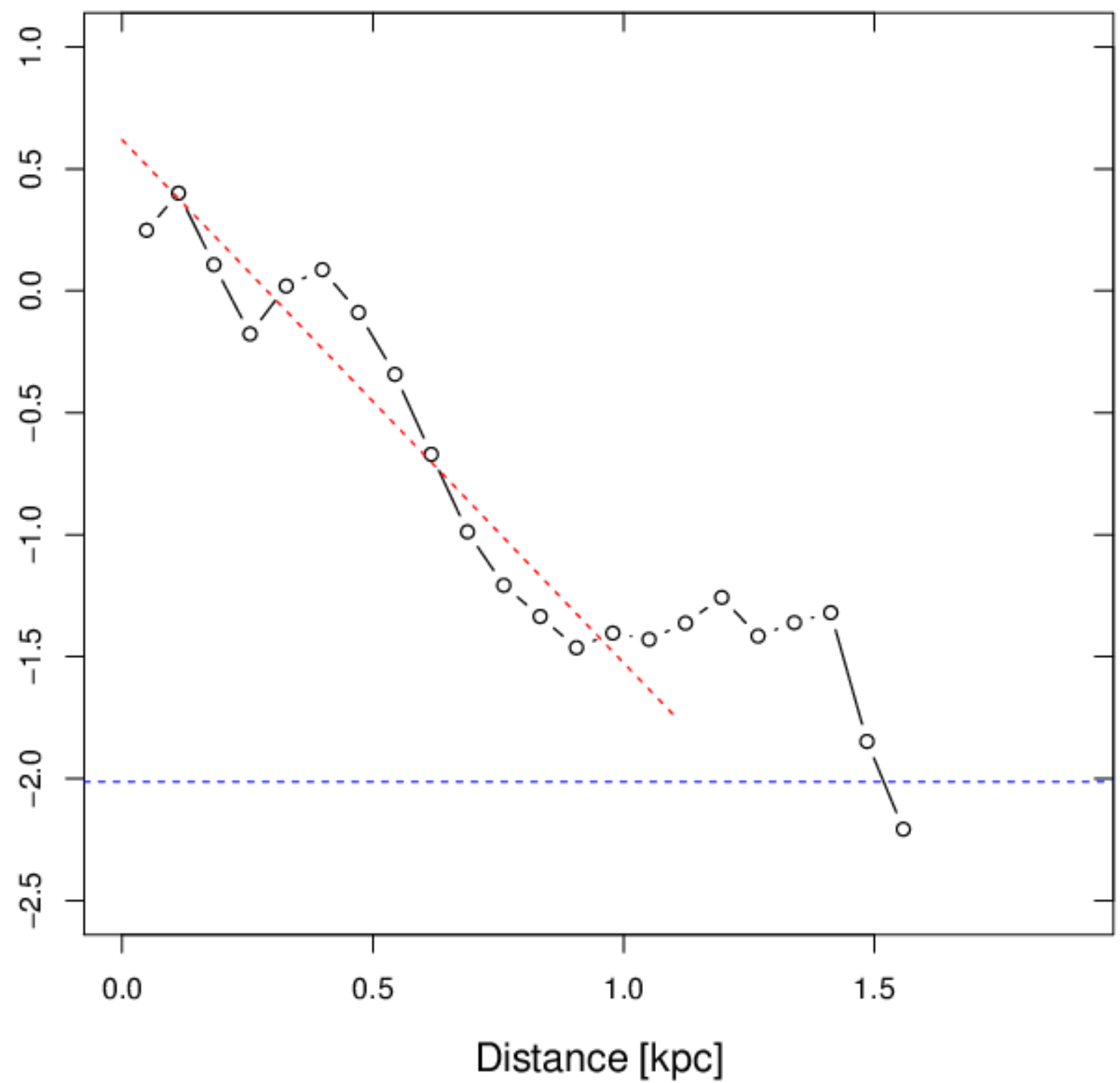}\\
\includegraphics[clip,width=.23\textwidth]{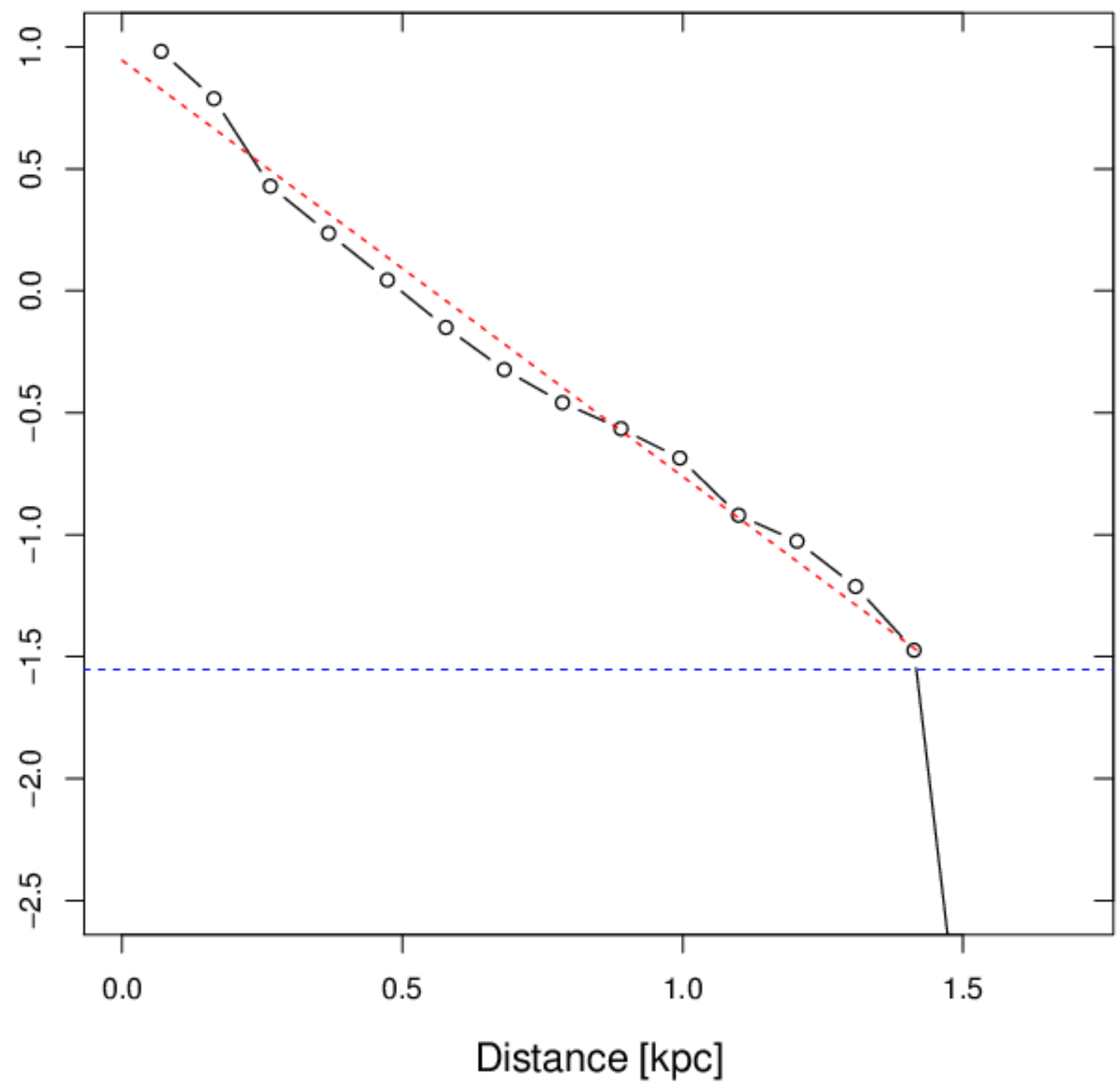} \includegraphics[clip,width=.23\textwidth]{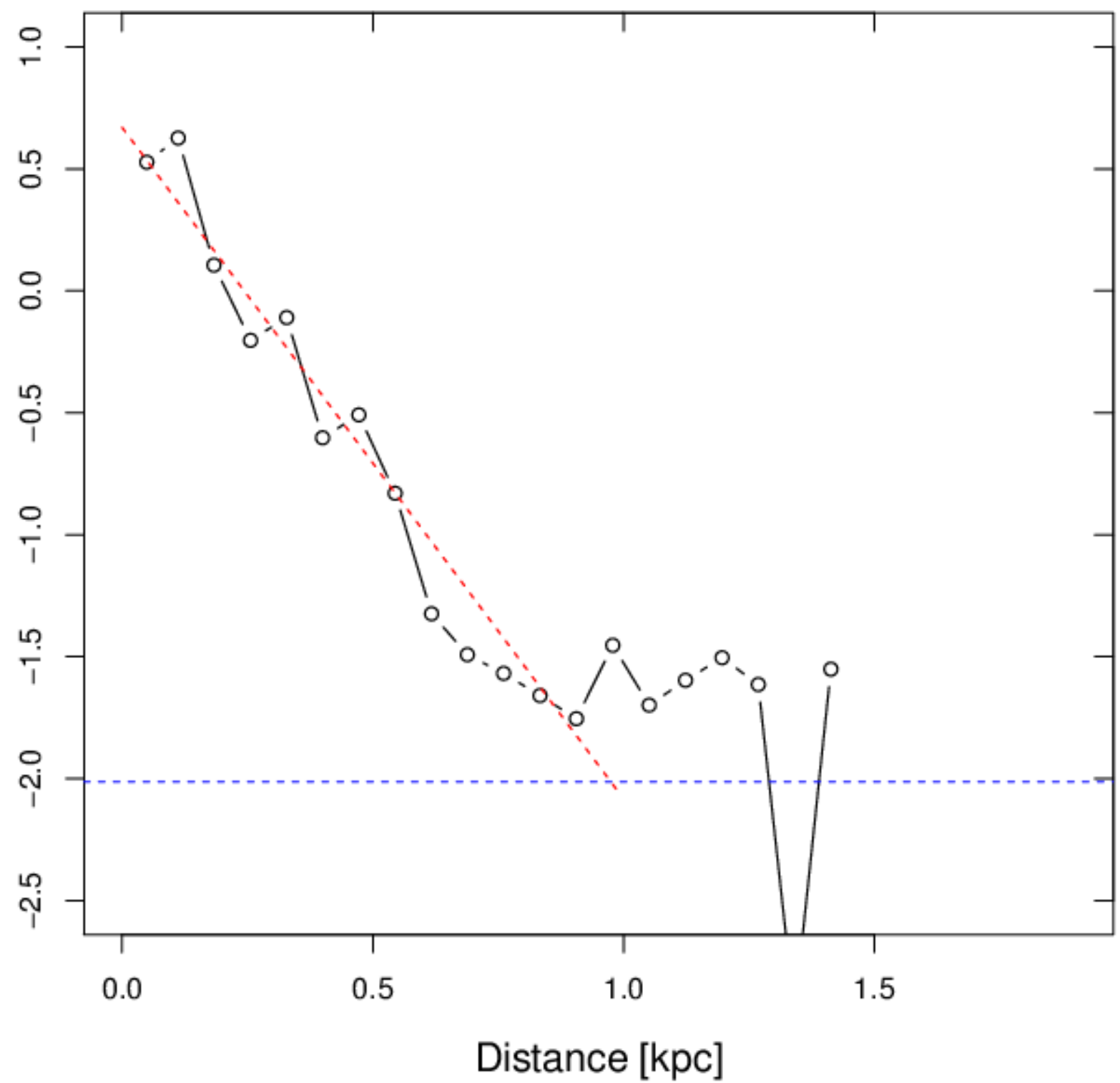}\\
\includegraphics[clip,width=.23\textwidth]{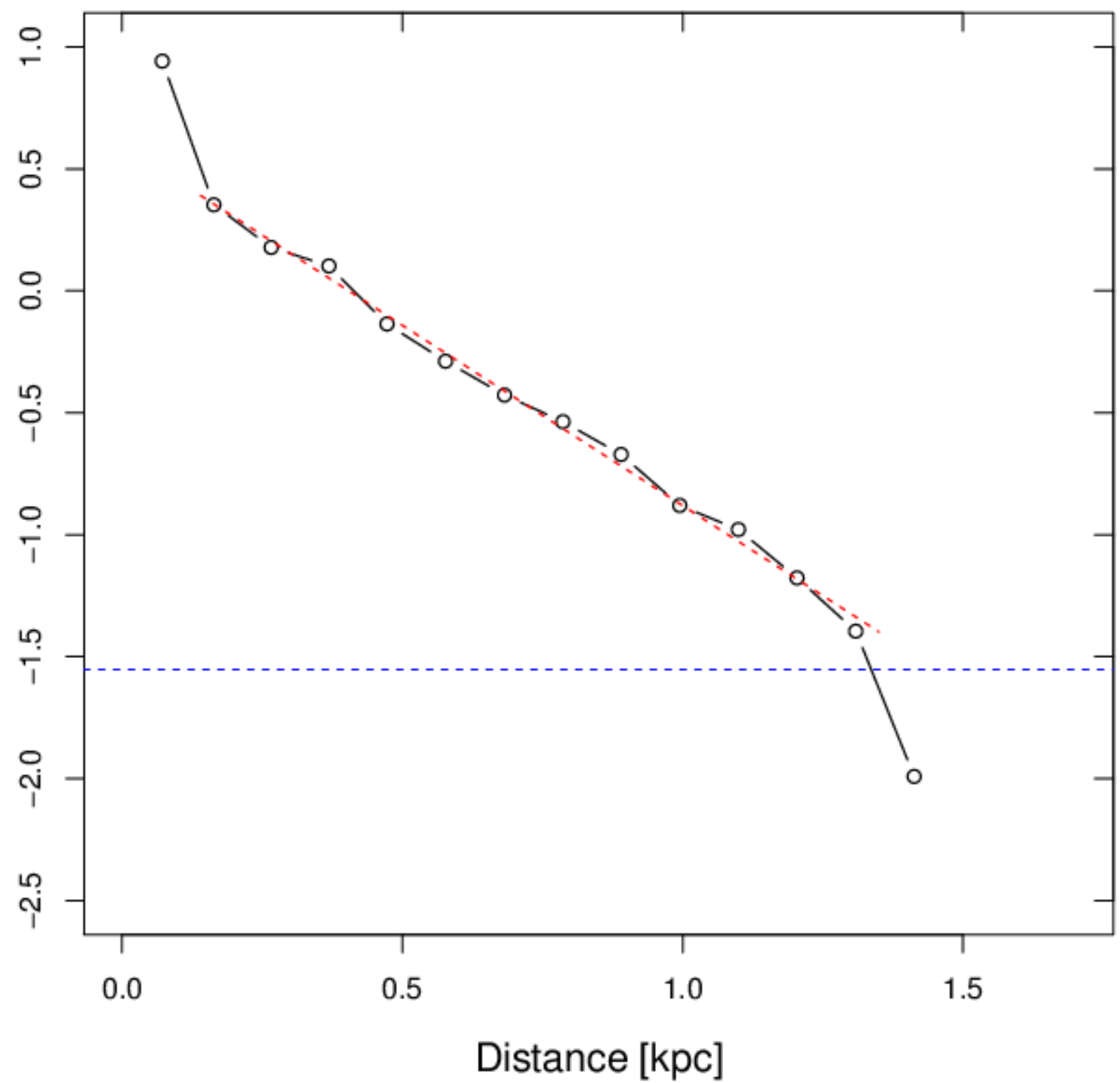} \includegraphics[clip,width=.23\textwidth]{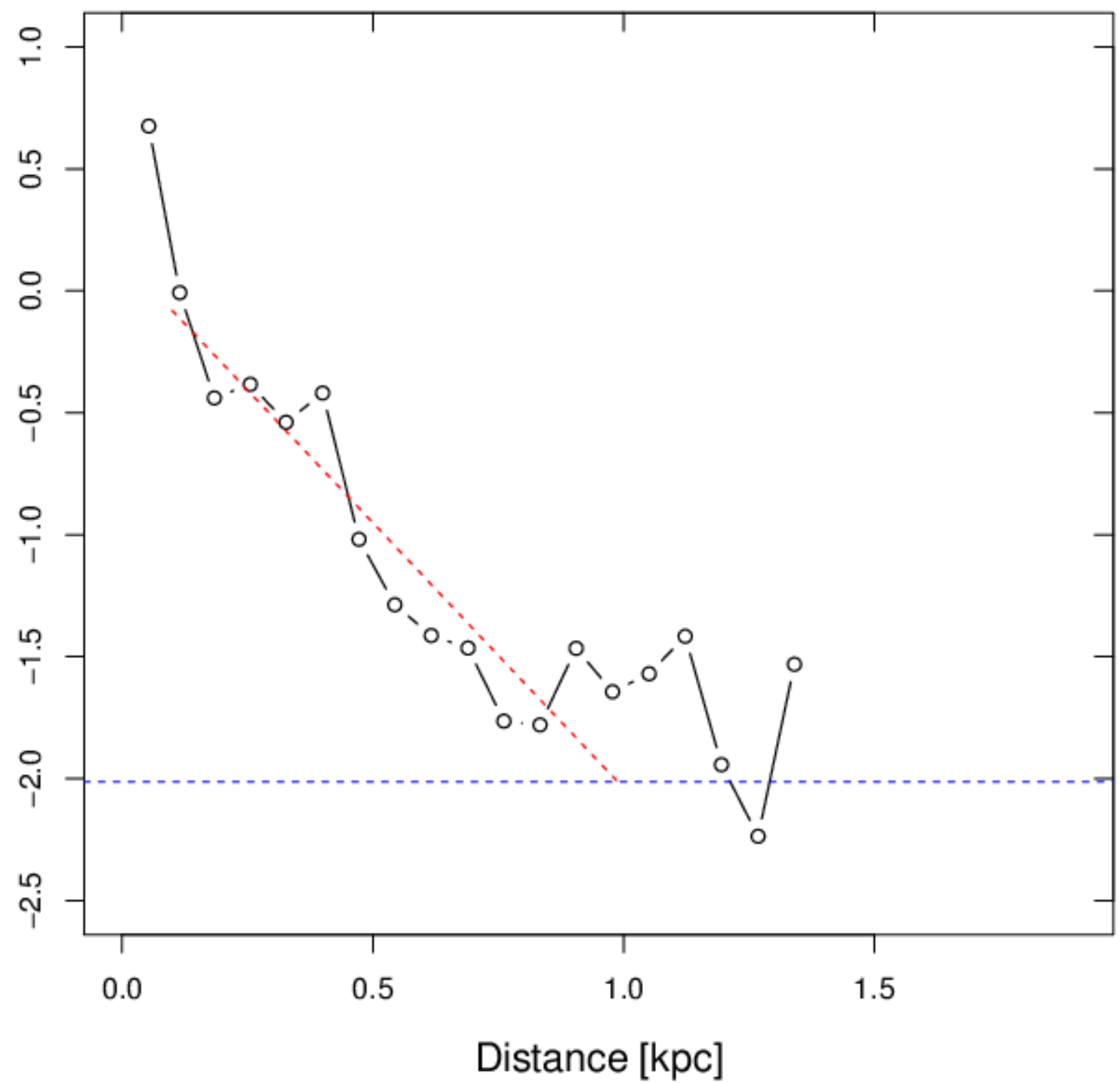}
\caption{Radio emission profiles of IC\,10 at $1.43$\,GHz (left) and $4.86$\,GHz (right)
with fitted exponential models (see text for details). The top panels correspond to all 
directions from the galactic center; the middle panels to directions within the 45\degr 
sector around $PA=35\degr$; the bottom panels to directions as in the middle ones, but 
with the tip of sectors moved to the northern giant \ion{H}{2} complex. The horizontal 
lines mark the rms noise levels of the radio maps.
}
\label{f:i10Slices21}
\end{figure}

\begin{table}[t]
\caption{Radio Emission Scale Length and Wind Velocities in IC\,10
}
\begin{center}
\begin{tabular}{ccccc}
\hline\hline
  Direction  &  Scale Length  & $\mathrm{B_{tot}}$ & $t_{syn}$ & $V_w$ \\
          &  (kpc)          & ($\mu G$)   & (Myr)       & (km\,s$^{-1}$) \\
\hline
           & $1.43$\,GHz & & \\
\hline
All        & $0.26$  & $13.5$ & $17.9$ & $26\pm4$ \\
NE ($PA=35\degr$)  & $0.25$  & $13.5$ & $17.9$ & $25\pm 4$ \\
NE from northern \ion{H}{2} complex & $0.29$ & $20.0$ &  $9.9$ & $52\pm 8$ \\
\hline
           & $4.86$\,GHz & & \\
\hline
All        & $0.19$  & $13.5$ & $9.7$ & $35\pm 5$ \\
NE ($PA=35\degr$) & $0.17$  & $13.5$ & $9.7$ & $30\pm 5$ \\
NE from northern \ion{H}{2} complex & $0.20$ & $20.0$ & $5.4$ & $66\pm 10$\\
\hline
\end{tabular}
\end{center}
\label{t:i10Hsyn}
\end{table}

The morphology of both the total power and PI, presented in Section
\ref{s:results}, suggests that  the magnetic field in IC\,10 is probably,
at least partially, shaped by a global galactic wind. The galactic halo that 
is well visible at $1.43$\,GHz map is not seen in any other ISM component. 
However, the expected hot gas in the halo (of $10^6$\,K) may exist but
is not detected in soft X-rays owing to possible large absorption
in the Milky Way (see Sect.~\ref{s:thermal}). Therefore, the radio emission
and magnetic field structure remain the main evidence for a galactic
wind in IC\,10.

The detection of magnetic fields far from star-forming regions indicates that 
also a population of CRs is delivered to distant regions in the halo of IC\,10. 
CRs can even drive thermal gas outflows with coupling facilitated by 
streaming instability (e.g.\ Dorfi \& Breitschwerdt \citeyear{dorfi12}).

Using the radio data, we will now attempt to estimate the bulk speed of the CR electrons,
which are assumed to be transported convectively from the disk into the halo. 
We use the approach pioneered by Heesen et~al.\ (\citeyear{heesen09}),
in which the bulk speed $V_{\rm w}$ of the wind can be approximated as the ratio of CR
electrons' scale length $l_\mathrm{e}$ to the CR electron lifetime. The value
of $l_\mathrm{e}$ can be derived from the synchrotron scale length ($l_\mathrm{syn}$)
in case of energy equipartition according to 
\begin{equation}
l_\mathrm{e} = \frac{3-\alpha_\mathrm{nth}}{2} l_\mathrm{syn}
\end{equation}
where $\alpha_\mathrm{nth}$ is the nonthermal spectral index. For IC\,10
we have $\alpha_\mathrm{nth}\approx -0.6$ (Chy\.zy et~al.\ \citeyear{chyzy03}).

The scale length $l_\mathrm{syn}$ is determined by fitting an exponential 
model to radial profiles, for which we integrated the radio emission in the sky plane
in radial rings. Because in IC\,10 the distributions of H$\alpha$ and \ion{H}{1} are very irregular
and galactic disk orientation unclear (Sect.~\ref{s:bfieldstructure}), we performed
the fitting in three cases. First, we integrated emission over all azimuthal 
angles in rings centered on the optical center of the galaxy. Second, we 
included emission only from a sector with an opening angle of $45\degr$ and a 
major-axis orientation at $PA=35\degr$. This sector is free of star-forming regions 
far from the disk, contrary to, e.g., the opposite (southeastern)
direction, which shows the chain of \ion{H}{2} regions (Sect.~\ref{s:results}).
In the third approach we made a similar fit but with the tip of the sector 
shifted to the center of the northern giant \ion{H}{2} complex. This case seems 
to be the most reasonable one to measure the advection of plasma as the 
\ion{H}{2} complex is probably the place of sources of CR acceleration and the 
magnetic field amplification process. The profiles of radio emission at 
$1.43$ and $4.86$\,GHz and results of model fitting are given in 
Fig.~\ref{f:i10Slices21} and Table \ref{t:i10Hsyn}. The obtained scale 
lengths for IC\,10 of about $0.3$\,kpc at $1.43$\,GHz are significantly smaller
than for massive spirals, e.g.\ NGC\,253, for which the scale is about 
$1.7$\,kpc at the same frequency (Heesen et~al.\ \citeyear{heesen09}).

For the estimation of the bulk speed of CR electrons we assume predominance of the
synchrotron losses ($t_\mathrm{e} = t_\mathrm{syn}$).  As stated earlier,
inverse Compton scattering is neglected in our analysis, so that the
actual electron lifetimes may be shorter. We also neglect adiabatic
losses, which could further shorten the electron lifetimes and enlarge the derived
velocity of the convective transport. The value of $t_\mathrm{syn}$ can be calculated
from the formula 
\begin{equation}
\frac{t_\mathrm{syn}}{\rm yr} = 8.352 \times 10^9 \left(\frac{E}{\rm
    GeV}\right)^{-1} \left(\frac{B}{\mu \rm G}\right)^{-2}
\end{equation}
where $B$ is the strength of the total magnetic field (see Sec. \ref{ss:IC10BfieldStrength})
and $E$ is the energy of the relativistic electrons derived at the frequency of observations
$\nu$:
\begin{equation}
\frac{E}{\rm GeV} = \left(\frac{\nu}{16.1\,\rm MHz}\right)^{0.5}\left(\frac{B}{\mu\rm G}\right)^{-0.5}.
\end{equation}
Taking the scale lengths determined for various cases and using the above 
formula, we computed the bulk speed of CR electrons $V_{\rm w}$ (Table \ref{t:i10Hsyn}).
The obtained values range from 25 to 66\,km\,s$^{-1}$.

We approximate the escape velocity $\mathrm{v_{esc}}$ from the disk of IC\,10 by the
maximum velocity of galactic rotation: $\mathrm{v_{esc}} \cong \sqrt{2}\, \mathrm{v_{max}}$.
Taking $\mathrm{v_{max}}$ of 30\,km\,s$^{-1}$ (Wilcots \& Miller \citeyear{wilcots98}), one
gets $\mathrm{v_{esc}}$ of 40\,km\,s$^{-1}$. Therefore, the estimated CR bulk 
speed in the vivid star-forming \ion{H}{2} complex of $59\pm 13$\,km\,s$^{-1}$ 
(a mean between $1.43$ and $4.86$\,GHz values) is considerably higher than the 
escape velocity. The average values of $V_{\rm w}$ calculated for the whole 
galaxy or for northeast direction are slightly smaller (25--35\,km\,s$^{-1}$)
than the escape velocity. These two cases assume weaker magnetic fields, which 
leads to less energetic outflows of plasma from the galaxy disk. However, even in 
these cases it is conceivable that the CR bulk speed exceeds the escape velocity, 
because of our conservative estimate of the CR energy losses. This corroborates 
our earlier suggestion that a magnetized galactic-scale wind exists in IC\,10.

A similar conclusion can be drawn from a rough estimate of the wind speed from the
radial slope of the spectral index profile (see Heesen et al.\ \citeyear{heesen09}). 
In the northeastern direction from the northern \ion{H}{2} complex the gradient 
of the spectral index between $1.43$ and $4.86$\,GHz is about 
$\Delta\alpha / \Delta r =-1.1\pm0.1$\,kpc$^{-1}$. For a magnetic field 
strength of $20\pm 3\,\mu$G we first derive the time derivative of the radio 
spectral index $\Delta \alpha / \Delta t$ and then obtain the bulk speed 
$\mathrm{V}=\Delta r / \Delta t =62\pm 17$\,km\,s$^{-1}$. This value agrees with the wind velocity
obtained from the radio profiles.

In our approach we assumed that plasma is transported convectively 
from the disk into the halo. We can now justify this assumption observing 
that the vertical profiles of the 1.43\,GHz radio continuum emission shown 
in Fig.\,\ref{f:i10Slices21} can be closely approximated by exponential 
functions, as expected for convection, whereas diffusion would lead to 
profiles resembling a Gaussian function (Heesen et al. \citeyear{heesen09}, 
\citeyear{heesen16}). This means that diffusion plays only a minor role, 
providing us with an upper limit of the diffusion coefficient as 
$D < (1-3)\times 10^{27}\,\mathrm{cm}^2 \,\mathrm{s}^{-1}$ (assuming $D=l_e^2/t_e$ 
and conditions given in Table \ref{t:i10Hsyn} for 1.43\,GHz). This value is 
approximately a factor of 10 smaller than the Milky Way value (Strong et al. \citeyear{strong07}), 
probably caused by the very turbulent magnetic field structure (Sect. \ref{ss:IC10BfieldStrength}).

Our estimate of the galactic outflow speed in IC\,10 is in agreement with the
outflow speeds in dwarf galaxies measured from the \ion{Na}{1} D absorption
line (below 100\,km\,s$^{-1}$), at least for those objects that have similar
circular velocities and SFRs (Veilleux et al.\ \citeyear{veilleux05}).
Thus, the observed extensive magnetic fields in the halo of IC\,10 are probably
not produced ``in situ'' by a dynamo process, but are likely ejected from
\ion{H}{2} complexes and dragged along with the galactic wind.

Galactic outflows are also considered as one of the mechanisms to form large-scale
X-shaped magnetic fields observed in edge-on spiral galaxies
(Sect.~\ref{s:bfieldstructure}). With  a velocity of about 60\,km\,s$^{-1}$ the 
CRs and magnetic fields in IC\,10, if advected with winds, can reach the most 
distant observable regions in the radio halo within $2 \times 10^7$\,yr. This 
corresponds well to the timescale of the currently ongoing starburst (Sect.~\ref{s:intro}).
The vertical extent of the radial (X-type) structure in IC\,10, when compared with
the optical disk size, is, relatively speaking, even larger than in edge-on spiral galaxies,
such as NGC\,4631 or NGC\,5775. In typical late-type spiral 
galaxies, the radio halo extends vertically only above star-forming regions, 
i.e. galaxies have ``sharp edges'' (Dahlem et al.\ \citeyear{dahlem06}). In IC\,10, however, 
the magnetic structure fills the entire, almost spherical radio halo 
(Fig.~\ref{f:i10_6_pi45}). Hydrodynamic simulations of CR-driven winds by Uhlig et
al.\ (\citeyear{uhlig12}) indeed predict that dwarf galaxies can experience
spherically symmetric outflows. To fully compare such models with IC\,10, the explicit
treatment of magnetic fields is necessary.

There is yet another impact of the proposed large-scale magnetized wind in IC\,10:
 galactic winds are considered as a potential process that enable galaxies
to magnetize the IGM. Dwarf galaxies, which have shallow gravitational potentials and
 burst-like star formation histories, could spread magnetic fields into areas very remote from the
regions of magnetic field and CR generation (Kronberg et~al.\ \citeyear{kronberg99};
Dubois \& Teyssier \citeyear{dubois10}; Chy\.zy et~al.\ \citeyear{chyzy11}). The 
radial topology of the ordered magnetic field in the halo of IC\,10, which currently 
undergoes a starburst (see Fig.~\ref{f:i10_BfieldDirections}), enables efficient 
transport of the magnetized plasma along the open field lines and nicely conforms
with this picture.

We trace the magnetic field in the halo of IC\,10 out to  a distance of about $1.4$\,kpc 
from the galaxy's center at $1.43$\,GHz. If the magnetized halo freely expands to a
stall radius of about 5\,kpc (see analytic modeling of Chy\.zy et~al.\
\citeyear{chyzy11}), we expect still strong fields there of $\approx$$0.5\,\mu$G.
Such extensive fields could be detected by probing very low energy CRs with low-frequency
observations (with WSRT, GMRT, or LOFAR), or by the method of RM
grids toward background polarized sources.  Such observations would allow
us to justify the role of starburst dwarf galaxies, such as IC\,10,
in the magnetization of the IGM. Local starbursting dwarf galaxies constitute 
the ideal laboratory for such studies, since they are, in the paradigm of the 
$\Lambda$CDM hierarchical structure formation, the closest analogs of the first 
galaxies in the Universe.

\section{Conclusions}
\label{s:conclusions}

We present a multifrequency radio continuum polarimetry study ($1.43$, $4.86$, 
and $8.46$\,GHz) of IC\,10, the nearest dwarf irregular galaxy in an ongoing starburst
phase. We use a combination of interferometric observations with the VLA and 
single-dish observations with the 100-m Effelsberg telescope. Our main 
conclusions are as follows:

\begin{enumerate}
\item Sensitive VLA observations at $1.43$\,GHz reveal a large radio envelope 
of IC\,10 in total intensity, extending up to $1.4$\,kpc away from the 
galaxy's center, thus almost two times further than the infrared-emitting 
galactic disk. The envelope is not seen at higher frequencies ($4.86$ and 
$8.46$\,GHz), suggesting substantial aging of CR electrons, actually confirmed 
by a gradual steepening of the  nonthermal spectral index with distance from the galactic disk.

\item  Compared to the total radio emission, the polarized emission of IC\,10 at
$4.86$ and $8.46$\,GHz is more clumpy, with the most pronounced polarized extension
located in the southern part of the galaxy, partially corresponding to the 
nonthermal superbubble. At $1.43$\,GHz, the polarized signal covers also a 
large fraction of the observed area around the galaxy, which we explain
as the Milky Way's foreground signal,  unrelated to IC\,10. Such a strong 
foreground polarized emission has not been found in any  other external galaxy before.

\item On the largest scales, the apparent B-vectors of the PI exhibit
a global ``X-shaped'' morphology of magnetic fields, with  a dominating radial 
component. The structure resembles the one observed in several nearby 
edge-on spiral galaxies with higher SFRs. No such structures have been 
observed in dwarf galaxies to date. The X-shaped magnetic structure can be 
caused by galactic winds pulling out magnetic fields, driven by stellar winds 
from young, massive stars and supernova explosions.

\item At the northern edge of the  nonthermal superbubble, the magnetic field 
orientation changes abruptly and is aligned with a prominent dust lane that 
can be seen in optical images. This reveals some external processes shaping 
the magnetic field, as infalling gas and/or tidal interactions,
resulting in compressed or stretched magnetic field  lines.

\item  In spite of the small size and mass of IC\,10, the magnetic field
strength has a mean of 14\,$\mu$G, similar to those found in massive
spiral galaxies. What sets the magnetic field structure apart is the
dominating random component with a degree of field order of only $0.17\pm 0.07$, so
that the ordered component has a mean strength of merely $2.3\pm 0.7\,\mu$G. Locally, 
the field is strongest in the southern and northern bright \ion{H}{2} 
complexes, where it reaches $29\,\mu$G and is almost entirely random. 
Farther away from the star-forming sites, the magnetic field becomes 
gradually more ordered, while at the same time the field strength drops. 
The nonthermal superbubble and the ``tangle region'' have field strengths 
of 22 and $15\,\mu$G, respectively. In the centers of \ion{H}{1} holes, 
the field  strength is about 7--$10\,\mu$G and of various degrees of field 
order ($0.1$--$0.5$). Even in the most distant galaxy outskirts (about 1.4\,kpc 
from the center), the magnetic field is still strong (with strength of 
$\approx$$7\,\mu$G) and partly ordered (with a degree
of $0.2$--$0.3$).

\item We detect Faraday rotation of about $-150\pm 69\,$rad\,m$^{-2}$ 
in the giant \ion{H}{2} complex, which corresponds to a weak regular magnetic field 
of $2\,\pm1\,\mu$G. We rule out a scenario for global operation of the mean-field 
dynamo process in IC\,10, as the calculated large-scale dynamo number is 
subcritical and the e-folding time for amplification of the regular field is 
relatively long ($>$$8\times10^8$\,yr, according to MHD simulations). Therefore, 
the strong total magnetic fields in IC\,10 probably arise from the small-scale dynamo.

\item The radio emission profiles at $1.43$\,GHz indicate a scale length
of the radio halo in IC\,10 of about $0.3$\,kpc, several times smaller than
found in massive spiral galaxies at the same frequency. In combination with 
the CR electron lifetime, this implies a CR bulk speed from regions 
dominated by a giant \ion{H}{2} complex of about 60\,km\,s$^{-1}$, exceeding 
the escape velocity. Thus, galaxy-wide magnetized winds can induce 
X-shaped magnetic fields and blow up the extensive radio halo visible
in total intensity at $1.43$\,GHz. Moreover, the magnetized plasma  escaping 
from the gravitational well of this dwarf irregular galaxy is able to seed
random and ordered magnetic fields in the IGM, in a similar fashion that 
posited for the first low-mass galaxies in the early Universe.

\end{enumerate}

The radio picture of IC\,10 is complex and manifests some magnetic patterns, 
which were found previously only in massive spiral galaxies. Therefore, this
dwarf irregular galaxy, one of our cosmic next-door neighbors that form the Local Group,
is a  valuable laboratory for our understanding of the generation and evolution of
magnetic fields in galaxies in general. Because of IC\,10's close proximity, it can
serve as one of the benchmarks at which realistic numerical MHD simulations of the ISM
and galactic outflows can be tested against. IC\,10 holds valuable clues as to how low-mass 
galaxies acquire gas for triggering starbursts and how they amplify
magnetic fields and eventually spread them out into the IGM. Having a clumpy
disk and  an intense star-forming activity, which provides
significant feedback  to the magnetized plasma, it  allows us to
gain precious insight in processes that regulate the formation of protogalaxies
in the early Universe.

\begin{acknowledgements}
We are thankful to Aritra Basu and the anonymous referee for very useful 
comments. This work was supported by the Polish National Science Center
grant No. 2011/03/B/ST9/01859 (observations) and 2012/07/B/ST9/04404 
(simulations and analysis). V.H. acknowledges support from the Science and 
Technology Facilities Council (STFC) under grant ST/J001600/1. R.B. and D.J.B. 
acknowledge support by the DFG Research Unit FOR1254. We acknowledge the use 
of the HyperLeda (http://leda.univ-lyon1.fr) and NED (http://nedwww.ipac.caltech.edu)
databases.
\end{acknowledgements}

\end{document}